\documentclass[iop, numberedappendix, onecolumn]{emulateapj}	
\usepackage{grffile}    
\usepackage{amsmath}
\usepackage{graphicx}
\usepackage{threeparttable}
\usepackage{times}
\usepackage{color}
\usepackage{verbatim}

\makeatletter

\usepackage{graphics}
\usepackage{grffile}
\usepackage{rotating}
\usepackage{natbib}

\newcommand{\be}{\begin{eqnarray}}
\newcommand{\ee}{\end{eqnarray}}
\newcommand{\etal}{et al.}

\def\gtrsim{\mathrel{\hbox{\rlap{\hbox{\lower4pt\hbox{$\sim$}}}\hbox{$>$}}}}
\def\lesssim{\mathrel{\hbox{\rlap{\hbox{\lower4pt\hbox{$\sim$}}}\hbox{$<$}}}}

\newcommand{\YX}[2]{#1_{\rm #2}}
\newcommand{\tDM}{\YX{t}{DM}}

\newcommand{\DM}{\rm DM}
\newcommand{\SM}{\rm SM}
\newcommand{\ds}{D_s}
\newcommand{\xvec}{{{\bf x}}}
\newcommand{\cnsq}{\rm C_n^2}
\newcommand{\xvecp}{{{\bf x}^{\prime}}}
\newcommand{\thetavec}{{ \mbox{\boldmath $\theta$} }}
\newcommand{\bvec}{{ \mbox{\boldmath $b$} }}
\newcommand{\dnud}{\Delta\nu_d}                 
\newcommand{\Dtiss}{\Delta t_{\rm ISS}}         
\newcommand{\taud}{\tau_d}                      
\newcommand{\epsvec}{{ \mbox{\boldmath $\epsilon$} }}

\newcommand{\Xarray}{{\bf X}}
\newcommand{\Carray}{{\bf C}}

\newcommand{\Dvec}{{\bf D}}

\newcommand{\Dp}{{D^{\prime}}}
\newcommand{\xpvec}{{\bf x^{\prime}}}

\newcommand{\zbar}{{ z}}
\newcommand{\dz}{\delta z}

\newcommand{\nup}{{\nu^{\,\prime}}}
\newcommand{\nul}{\nup}			
\newcommand{\nuh}{\nu}				

\newcommand{\qvec}{{\bf q}}
\newcommand{\qvecperp}{{\bf q}_\perp}
\newcommand{\qperpvec}{\qvecperp}
\newcommand{\qperp}{{q_\perp}}

\newcommand{\xpsrvec}{{{\bf x}_{p}}}
\newcommand{\xevec}{{{\bf x}_{e}}}
\newcommand{\vpsrvec}{{{\bf v}_p}}
\newcommand{\vevec}{{{\bf v}_e}}

\newcommand{\Pne}{P_{\delta n_e}}
\newcommand{\nebar}{{\overline{n}_e}}

\newcommand{\DMbar}{\overline{\DM}}

\newcommand{\Phibar}{\overline{\Phi}}

\newcommand{\xvecpp}{{\bf x^{\prime\prime}}}

\newcommand{\Anuxz} {{A_\nu(\xvec,z)}}

\newcommand{\Anuxpz} {{A_{\nuh}(\xvecp, \zp)}}
\newcommand{\Anuxppz } {{A_{\nul}(\xvecpp, \zpp)}}

\newcommand{\Anuxxpz} {{A_{\nuh}(\xvec-\xvecp, \zp)}}

\newcommand{\zp}{z^{\,\prime}}
\newcommand{\zpp}{z^{\,\prime\prime}}

\newcommand{\Cdmbar}{{C_{\delta\DMbar}}}
\newcommand{\Cdmbarnn}{\Cdmbar(\nuh, \nuh)}
\newcommand{\Cdmbarnpnp}{\Cdmbar(\nul, \nul)}
\newcommand{\Cdmbarnnp}{\Cdmbar(\nul, \nuh)}
\newcommand{\xbar}{{\overline{x}}}
\newcommand{\xbarvec}{{\overline{\xvec}}}
\newcommand{\dxvec}{{\delta \xvec}}

\newcommand{\DMbarrms}{\DMbar_{\rm rms}}
\newcommand{\Phibarrms}{\Phibar_{\rm rms}}

\newcommand{\sigmadDMbar}{\sigma_{\footnotesize \DMbar}}

\newcommand{\Deff}{D_{\rm eff}}
\newcommand{\SMeff}{\SM_{\rm eff}}
\newcommand{\sigtheta}{\sigma_{\theta}}
\newcommand{\sigthetas}{\sigma_{\theta_s}}
\newcommand{\sigDM}{\sigma_{\DM}}

\newcommand{\sigxnu}{\sigma_X(z,\nu)}
\newcommand{\sigxnul}{\sigma_X(z,\nul)}
\newcommand{\sigxnuh}{\sigma_X(z,\nuh)}
\newcommand{\sigxnulsq}{\sigma^2_X(z,\nul)}
\newcommand{\sigxnuhsq}{\sigma^2_X(z,\nuh)}

\newcommand{\sigxnuhbetatwo}{\sigma_X^{\beta-2}(z,\nuh)}

\newcommand{\xtheta}{x_{\theta}}

\newcommand{\KDM}{K}
\newcommand{\nerms}{{n_e}_{_{\rm rms}}}

\newcommand{\tinfty}{t_{\infty}}
\newcommand{\tinftyhat}{\widehat{t}_{\infty}}
\newcommand{\FWHM}{\theta_{\rm FWHM}}
\newcommand{\dunit}{2\pi}		

\newcommand{\sigDMcoeffSM}{3.76 \times10^{-5}\, {\rm pc~cm^{-3}}}
\newcommand{\sigDMcoeffPhiF}{4.42 \times10^{-5}\, {\rm pc~cm^{-3}}}
\newcommand{\sigTOAcoeffSM}{156}		
\newcommand{\sigTOAcoeffPhiF}{184}		
\newcommand{\dDMone}{3.84 \times10^{-8}\, {\rm pc~cm^{-3}}}
\newcommand{\Dne}{D_{\delta n_e}}

\newcommand{\pasa}{PASA}

\begin{document}

\vspace{-0.1in}
\title{
Frequency-Dependent Dispersion Measures and Implications for Pulsar Timing  
}
 \author{
J. M. Cordes\altaffilmark{1},
R. M. Shannon\altaffilmark{2}, 
 D. R. Stinebring\altaffilmark{3} 
\\
}
\altaffiltext{1}{Astronomy Department, Cornell University, Ithaca, NY 14853, USA; cordes@astro.cornell.edu}
\altaffiltext{2}{CSIRO Astronomy and Space Science, Box 76, Epping, NSW 1710, Australia;ryan.shannon@csiro.au}
\altaffiltext{3}{Department of Physics \& Astronomy, Oberlin College,Oberlin, OH 44074, USA; dan.stinebring@oberlin.edu}



\begin{abstract}
We analyze the frequency dependence of the dispersion measure (DM), the column density of free electrons to a pulsar,  caused by  multipath scattering from small scale electron-density fluctuations.  The DM is slightly different along each propagation path and the transverse spread of paths varies greatly with frequency, yielding arrival time perturbations that scale differently than the inverse square of the frequency, the expected dependence for a cold, unmagnetized plasma.    We quantify DM and pulse-arrival-time perturbations analytically for  thin phase screens and extended media and verify the results with simulations of thin screens.  The rms difference between DMs across an octave band near 1.5~GHz
$\sim 4\times10^{-5}\,{\rm pc\ cm^{-3}}$ for pulsars at  $\sim 1$~kpc distance.
Time-of-arrival errors resulting from chromatic DMs are of order a few  to hundreds of nanoseconds for pulsars 
with DM $\lesssim 30$~pc~cm$^{-3}$  observed across an octave band but increase rapidly to microseconds or larger for larger DMs and wider frequency ranges.
Frequency-dependent DMs introduce correlated noise into timing residuals whose power spectrum is `low pass' in form.  
The correlation time is of order the geometric mean of the refraction times for the highest and lowest radio frequencies used
and thus ranges from  days to years, depending on the pulsar.  We discuss the implications for methodologies that use large frequency separations or wide bandwidth receivers for  timing measurements.  Chromatic DMs are partially mitigable by using an additional chromatic term in  arrival time models. 
Without mitigation, our results provide an additional term in the noise model for pulsar timing; they also indicate
that in combination with measurement errors from radiometer noise, an arbitrary increase in total frequency range (or bandwidth) will yield diminishing benefits and may be detrimental to overall timing precision. 
\end{abstract}



\section{Introduction}

Pulsar arrival times with sub-microsecond accuracy  are required for the  detection of   nanohertz-frequency  gravitational waves using pulsar timing arrays \citep[e.g.][]{1990ApJ...361..300F}  and have payoffs in related  areas, such
as precision tests of theories of gravity \cite[][]{2006Sci...314...97K}, determination of neutron star masses \cite[][]{2013Sci...340..448A}, and characterization of microstructure in
the interstellar medium (ISM)  \cite[][]{1977ApJ...214..214I}. 

A pulse's time of arrival (TOA)  includes a group delay term, 
$ \tDM(\nu) = \KDM \nu^{-2} \DM$, where the dispersion measure (DM) is the integral of
the electron number density along the line of sight, $\nu$ is the frequency  and the constants are $\KDM = c r_e / 2\pi$ and the classical electron radius  $r_e$.  A key step in arrival-time analysis is the removal of the dispersion
term.
Estimates of \DM\   based on TOA differences between two or more frequencies
show that \DM\ is epoch dependent for most well-studied pulsars \citep[][]{1977ApJ...214..214I, 1985MNRAS.214P...5H, 1990ApJ...349..245C, 1991ApJ...382L..27P, b+93, 1994ApJ...428..713K, 2005AstL...31...30I, 2006ApJ...645..303R, 2013ApJ...762...94D, 2013MNRAS.429.2161K, 2013MNRAS.435.1610P, 2014ApJ...787...82F}.   Temporal variations include
a slow, systematic trend as a pulsar  moves toward or away from us along with stochastic variations 
from pulsar motion that causes the line of sight to sample electron-density fluctuations on a variety of scales. 

 In this paper we  show that the DM is also frequency dependent   because the ISM is sampled differently due  to multipath scattering.  The DM along each ray path is slightly  different and the size of the ray-path bundle increases monotonically with decreasing frequency, leading to a net difference in dispersive delays in the sum over all ray paths.  The variation with frequency of  $\DM(\nu, t)$ stays constant for a  refraction time, which is the time scale for the ray-path bundle  to move
 by its transverse extent and  can range from hours to years \citep[][]{rcb84, ssh+00}.   
Our analysis includes the general case of an arbitrary medium described by a wavenumber spectrum that varies along the line of sight (LOS).  We give specific results for the case of a thin screen and for a medium with homogeneous statistics for electron-density variations.  We compare analytical results with simulations of
scattering and dispersion.   Our analysis is for the strong scattering regime where all of the flux from a pulsar is scattered.   In the summary and conclusions section we briefly discuss the weak scattering regime and its role in precision timing. 

Our results complement those of \citet[][]{2015ApJ...801..130L}, who quantified TOA errors that result from
estimates for DM that make use of timing observations at different frequencies that are not made on the same day.
Here we show that even simultaneous multi-frequency measurements yield TOA errors because the DM varies
with frequency. 

Section~\ref{sec:DMtf} derives the basic effect and gives specific results for a thin screen and for a medium with uniform statistics. 
Section~\ref{sec:timing} concerns implications for pulsar timing methodology and precision. 
Section~\ref{sec:conclusions} discusses our results and summarizes our conclusions. 
In Appendix~\ref{app:scatt} we derive scattering quantities  and 
in Appendix~\ref{app:2fDMSF} we derive the RMS two-frequency variation  of DM. 

\begin{figure}[h!]
\begin{center}
\includegraphics[scale=0.42]{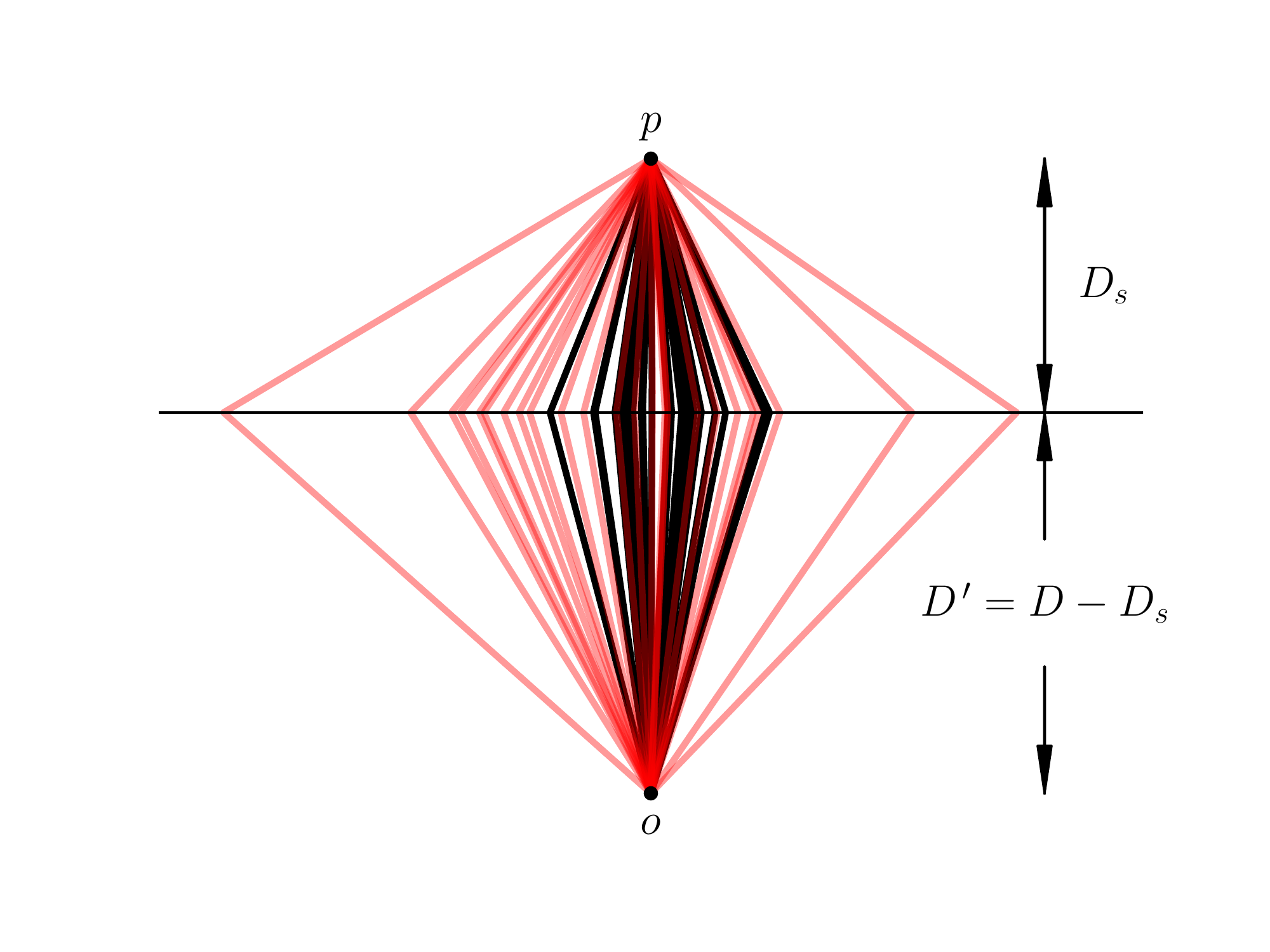}
\hspace {-0.2in}
\includegraphics[scale=0.42]{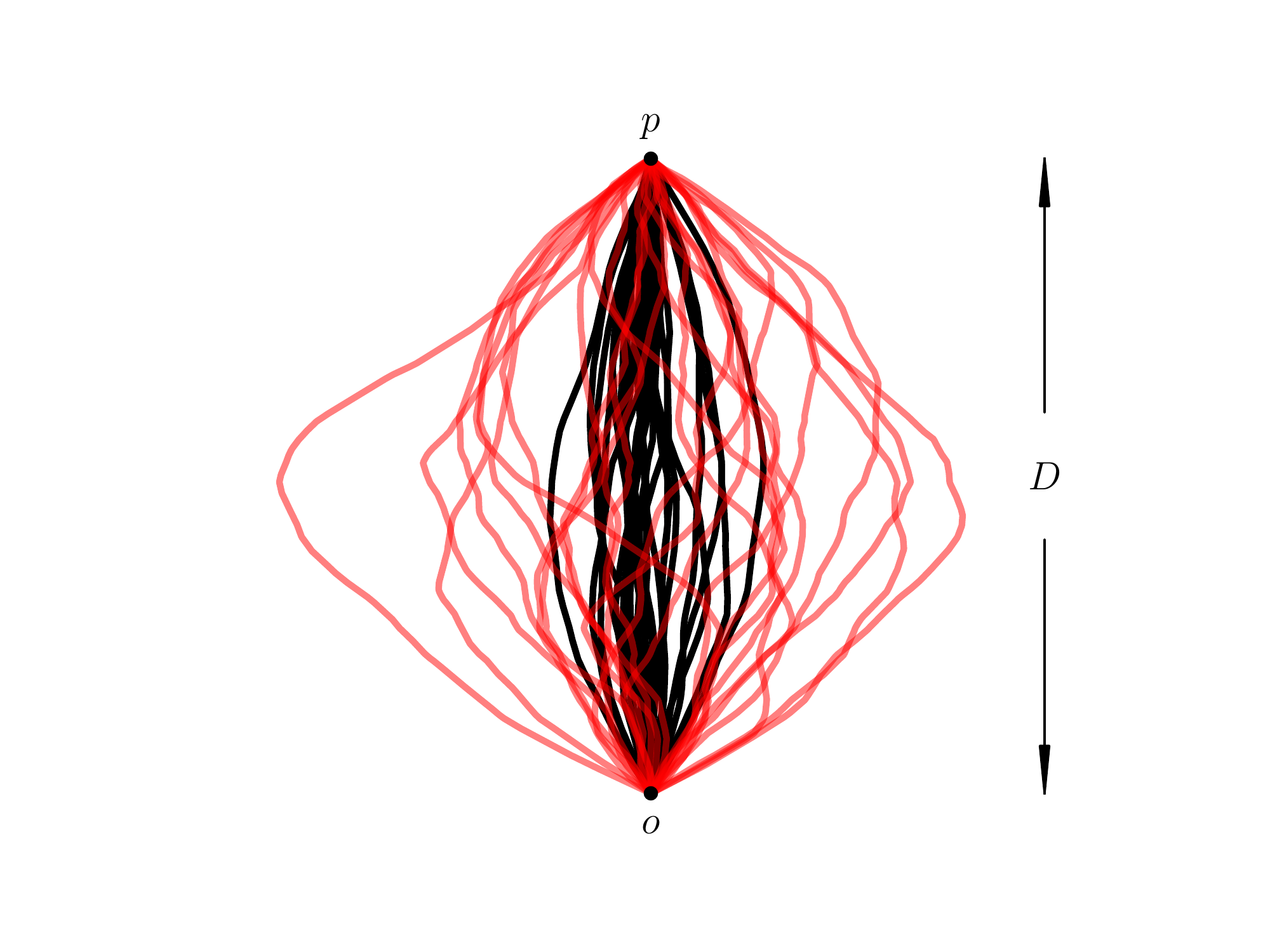}
\caption{\footnotesize 
Geometries for scattering from a thin screen (left) and filled medium (right). 
Black lines show simulated ray paths at frequency $\nu$  and red lines are for frequency $\nup  = \nu/2$. 
The pulsar-observer distance is $D$.  
 The thin screen is at a distance $\ds$ from the pulsar at $p$ and $\Dp = D-\ds$ from the  observer at $o$.   
\label{fig:raypaths}
}
\end{center}
\end{figure}

\vspace{0.1in}

\section{Dispersion Measure Variations in Time and Frequency}
\label{sec:DMtf}


There are several underlying causes for  temporal variations of DM  but only electron density fluctuations in the ISM produce  a large enough frequency dependence to be important in precision timing \footnote{The effects of the solar wind can be observed in millisecond pulsar observations within $\sim 10$-solar radii of the Sun  \cite[][]{2007ApJ...671..907Y}; however these observations are unsuitable for pulsar timing because of elevated system temperatures and instrumental distortions associated with the radiation from the nearby Sun.}.   
For a spatially uniform $n_e$,  the \DM\  is independent of frequency  and any epoch dependence
 comes from  changes in the pulsar distance $D$.   Motions of the pulsar and Earth with a combined velocity $100V_{100}$~km~s$^{-1}$  parallel to the LOS yield  $\delta\DM \sim 3\times10^{-6}~{\rm pc~cm^{-3}} V_{100} t_{\rm yr} n_{e_{0.03}}$ over $t_{\rm yr}$ years for a typical average electron density $n_e = 0.03$~cm$^{-3}$.
Linear trends in DM are indeed seen \citep[e.g.][]{2013MNRAS.429.2161K} but this effect is likely to be frequency
independent. 
 
Chromatic DMs result from multipath propagation caused by diffraction and refraction 
from interstellar electron density variations.    The effect we identify is not due to geometrical path-length differences alone.  Gravitational lensing, for example, could produce
multiple ray paths through a medium with constant electron density,  but the resulting DM variations  would be negligible ($\lesssim 10^{-16}$~pc~cm$^{-3}$) given observational bounds on time delays between paths 
$< 1~\mu s$.  Moreover, they would be achromatic.  

Density variations exist over a wide range of length scales from kpc to around $10^3$~km \citep[e.g.][]{ars95}
and  the smaller ones are responsible for multipath propagation.  The DM varies slightly between ray paths and    the cross-sectional area of the ray-path  bundle at any position along the LOS  is strongly frequency dependent, $\propto \nu^4$.    Density microstructure therefore plays two roles: causing multipath propagation and providing variable path integrals of the electron density.
Figure~\ref{fig:raypaths} shows frequency-dependent ray paths for  a thin scattering screen and for a medium that fills the volume between us and a pulsar; it also defines some of our notation.


We model the electron density to be $n_e(\xvec) = \overline{n}_e(\xvec) + \delta n_e(\xvec)$, 
where $\overline{n}_e(\xvec)$ is a constant local mean and $\delta n_e(\xvec)$ is 
 the zero-mean fluctuating part described by a wavenumber
spectrum $\Pne(\qvec, z)$ that can vary slowly along the  line of sight ($z$-axis). 
We adopt a power-law wavenumber spectrum  of the form
\be
\Pne(\qvec, z) = \cnsq(z) q^{-\beta}, \quad\quad q_0 \le q \le q_1.
\label{eq:pne}
\ee
The dependence on only the magnitude of $\qvec$ is a simplifying assumption, discussed further in the
next subsection, that conflicts with
some observations that show elliptical scattered images but is consistent with others that show
circular images.

The  electromagnetic phase perturbation $\phi$ from the index of refraction in a cold  plasma\footnote{We assume the electron density and magnetic field are small and that frequencies are large enough so that only the term 
linear in electron density is important.}  is proportional to the integrated electron density along a ray path \citep[][Appendix A, Eq. A5]{r90}, 
\be
\phi(\xvec) = -\lambda r_e \int_{\rm _{ray}} \!\!\!\!\! dz\, n_e(\xvecp(z), z),
\label{eq:phi_ne}
\ee
where $\xvec$ is the  transverse location in the observation plane a distance $D$ from the pulsar 
while $\xvecp$ is a transverse vector at a location $z$ along the LOS.

For any single ray path the phase perturbation corresponds to a dispersive time delay
$\tDM = (1/2\pi)(d\phi /d\nu)$ characterized by a dispersion measure,  $\DM = -\phi/ \lambda r_e$. 
Converting $\lambda r_e$ to DM units at 1~GHz yields a natural value for one radian of phase, 
$\delta \DM_1 = 3.84 \times 10^{-8}$~pc~cm$^{-3}$.   Since we know that pulsars are scattered significantly
at radio frequencies, phase perturbations $\phi_F$ on the Fresnel scale $r_F = \sqrt{\lambda D/2\pi} \sim 10^{11}$~cm are necessarily many radians. 
The Fresnel scale corresponds to less than one day of transverse motion for  pulsar velocities $\sim 100$~km~s$^{-1}$. We therefore  expect DM to vary  by large multiples of 
$\delta \DM_1$  on time scales of weeks and longer because the RMS phase grows with increasing time span
for media with a spectrum like that in Eq.~\ref{eq:pne}.   By similar reasoning, we expect 
frequency-dependent DM differences $\gg \delta \DM_1$ 
because ray paths at  widely spaced frequencies  have separations much larger than  the Fresnel scale. 

\begin{figure}[t!]
\begin{center}
\includegraphics[scale=0.5]
{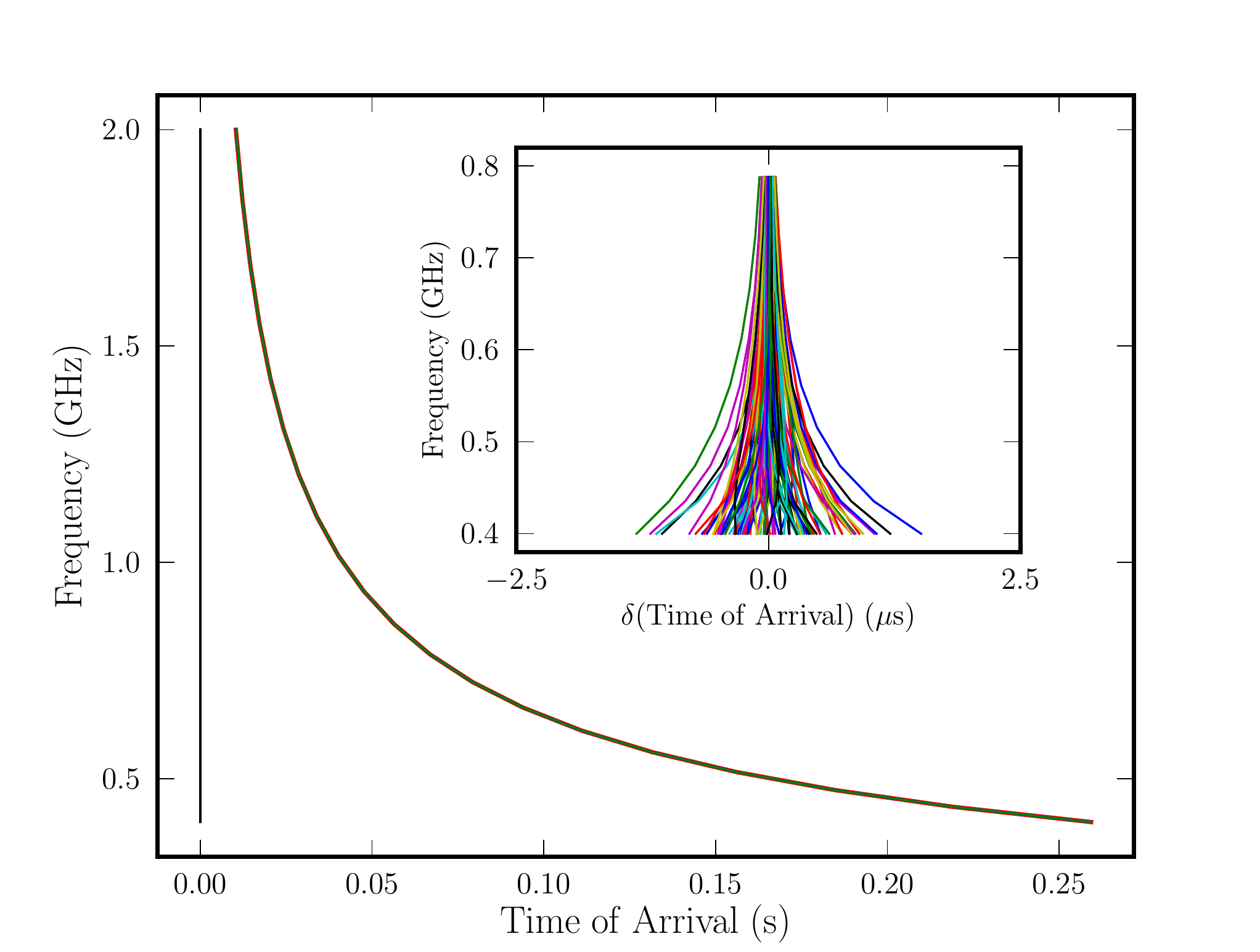}
\caption{ \footnotesize
Trajectories of a dispersed pulse in the time-frequency plane for 100 realizations using a phase screen with 
$\phi_F = 30$~rad at 1 GHz.  The dedispersed trajectories are shown aligned with $t=0$,
and the inset shows details of the dedispersed trajectories for the lower part of the frequency band.
\label{fig:DMsweep}
}
\end{center}
\end{figure}

Figure~\ref{fig:DMsweep} shows the manifestation of chromatic DMs in simulated arrival times vs. frequency.  The curve in the main part of the figure is dominated by the mean  $\DM = 10$~pc~cm$^{-3}$.  The inset shows
the deviations from the mean curve from 0.4 to 0.8 GHz  for 100 realizations of a phase screen with $\phi_F = 30$~rad, demonstrating
the spread in arrival times over $\pm 1~\mu s$ at the lowest frequency. 

\subsection{Simulations of Phase Screens and Ray-path Averaging}
\label{sec:sims}

 \begin{figure}[t!]
\begin{center}
\includegraphics[scale=0.80]
{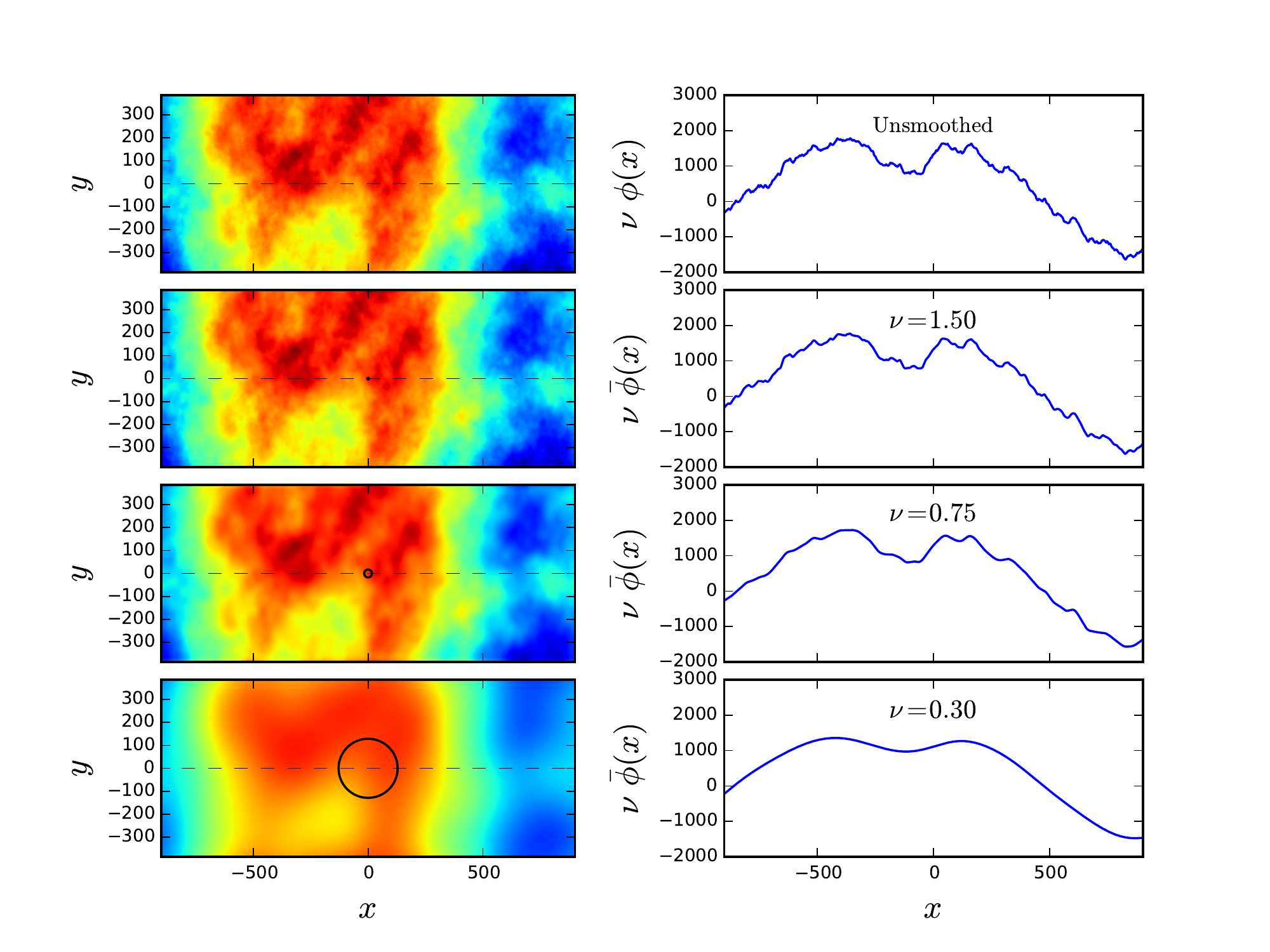}	
\caption{\footnotesize  
Phase screen for electron-density fluctuations with a Kolmogorov wavenumber spectrum that has
$\phi_F = 5$~rad RMS phase variation on the Fresnel scale.     The $x$ and $y$ coordinates
are in units of the Fresnel scale at 1~GHz and the RMS  phase difference grows as $\Delta^{5/6}$ between two points separated by $\Delta = [ (\Delta x)^2 + (\Delta y)^2]^{1/2}$.    For a velocity $V_x = 100$~km~s$^{-1}$, each sample corresponds to 0.28~d and the entire span of $x$ corresponds to 2~yr. 
Left: phase screens without (top) and with spatial smoothing.  The bottom three panels are for frequencies
of 1.5, 0.75, and 0.3~GHz, as labeled in the right hand panels.
Plotted circles (barely discernible in the middle two frames) are  projected scattering-disk sizes    and represent the region with $1/e$ radius on the phase screen that is averaged 
to produce the DM difference at any given epoch and frequency. The area scales as $\nu^{22/5}$.  
Right: phase variations corresponding to trajectories along the dashed lines in the left-hand panels.
The  1D curves are not smoothed versions of $\phi(x)$ in the top panel because the  smoothing is two dimensional.
\label{fig:bullseye}
}
\end{center}
\end{figure}

Phase screens  were simulated  using approaches similar to those presented in 
\citet[][]{1986ApJ...310..737C, 1987ApJ...315..666C, fc90,1991ApJ...366L..33H}
and \citet[][]{2010ApJ...717.1206C}.
An array in the wavenumber domain was filled with white, Gaussian, Hermitian noise
 shaped by the square root of the  ensemble-average power spectrum.  A 2D inverse FFT yields the phase screen.  To include low-frequency components excluded by the FFT, we separately
 added  wavenumber components with periods up to three times the array size in each
 direction.  Phase fluctuations were scaled to a specified Fresnel phase at a fiducial frequency, which we took
 to be 1~GHz.   We exploited the scale invariance of the phase fluctuations by simulating a set of phase screens
 for small values $\phi_F = 5$~rad, which require fairly small 2D arrays,  and rescaling to larger  values.   We verified that this procedure was correct by explicitly simulating a few large phase screens.

 Figure~\ref{fig:bullseye} shows an example phase screen and the averaging areas projected onto the screen
 for a 5:1 range of frequencies.  DM differences are due solely to the change in averaging area.   
 At different epochs the averaging areas sample different parts of the screen, producing stochastic variations in DM that  are correlated over the time it takes for the averaging area to move by its diameter.   The difference in DM 
 between a pair of frequencies is constant over the correlation time for the higher of the two frequencies, 
 which has the smaller scattering disk.   
 
\begin{figure}[t!]
\begin{center}
\includegraphics[scale=0.41]
{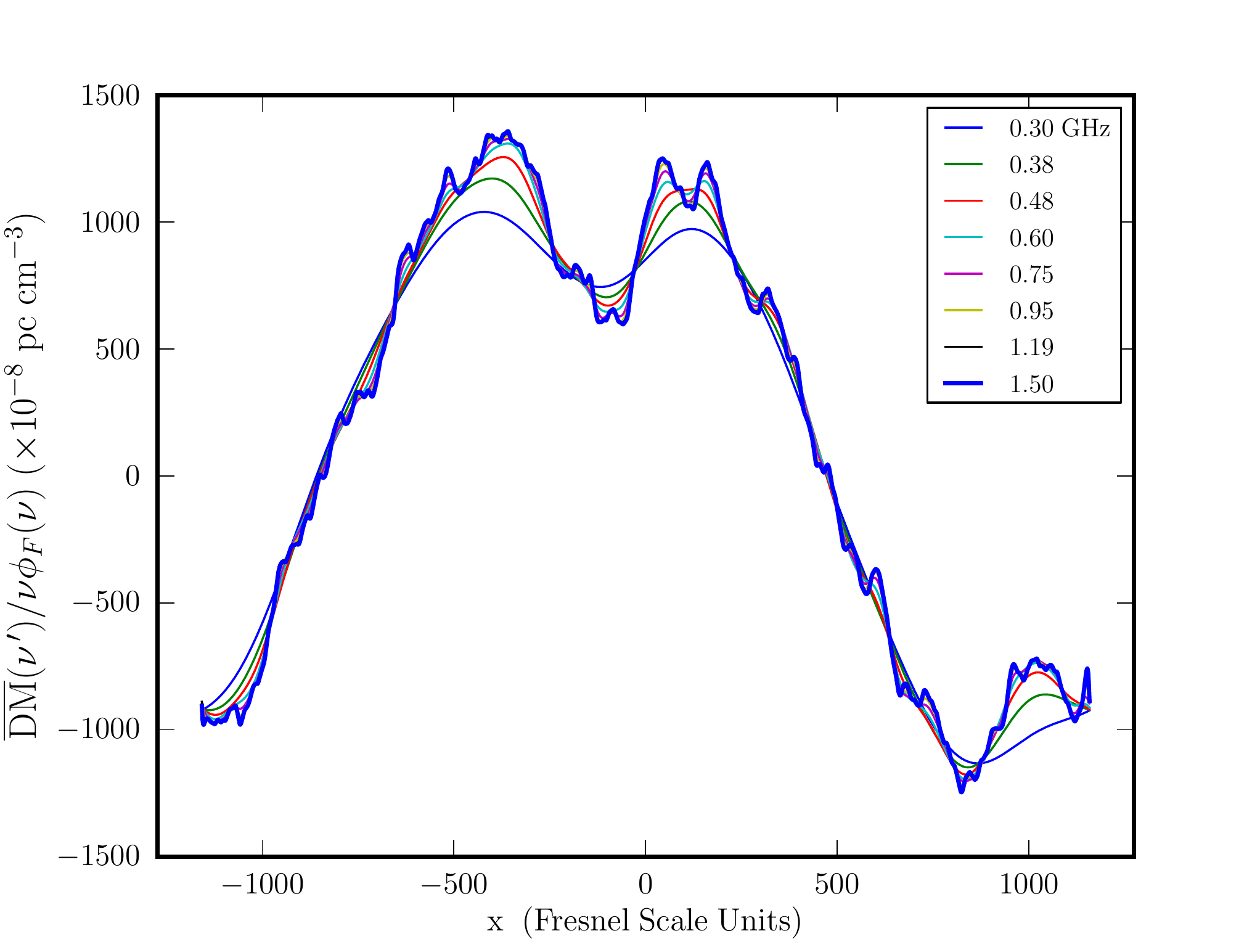}
\hspace{-0.3in}
\includegraphics[scale=0.41]
{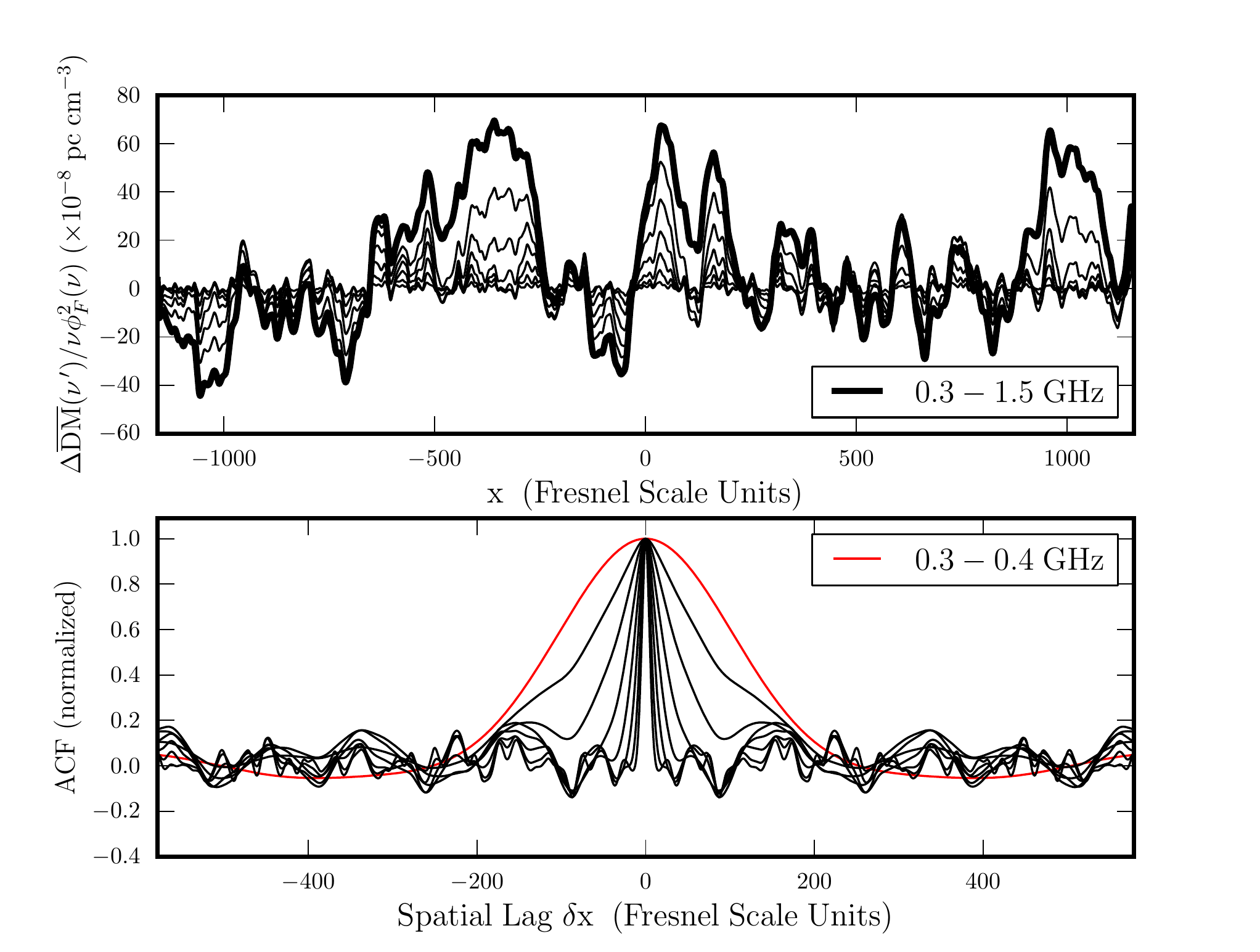}
\caption{\footnotesize  
(Left) A single realization showing  $\DM(t)$ at 8 frequencies from 0.3~GHz (red) to 1.5~GHz.  
 The DM values have been normalized by the quantity $\nu_1 \phi_F(\nu_1)$ for
$\nu_1 = 1$~GHz. (Right) Top panel: time series of $\DM(t)$ differences between  1.5 GHz  and all lower frequencies. The thinnest lines are for the frequencies nearest 1.5~GHz and the thickest for 0.3~GHz.
Bottom panel:
normalized autocorrelation functions of the  difference time series in the top panel along with the ACF of the DM 
difference between the two lowest frequencies (0.4 and 0.5~GHz, red).  The main lobe of each ACF near
zero lag represent the characteristic spatial scale of the DM difference.  Spatial units can be converted to
time units using $\Delta x = V_x \Delta t$.   Using 100~km~s$^{-1}$ velocity and a Fresnel scale 
for 1~GHz and 1~kpc distance, one spatial unit corresponds to 0.14~d. 
\label{fig:DMtnu}
}
\end{center}
\end{figure}

\begin{figure}[t!]
\begin{center}
\includegraphics[scale=0.41]
{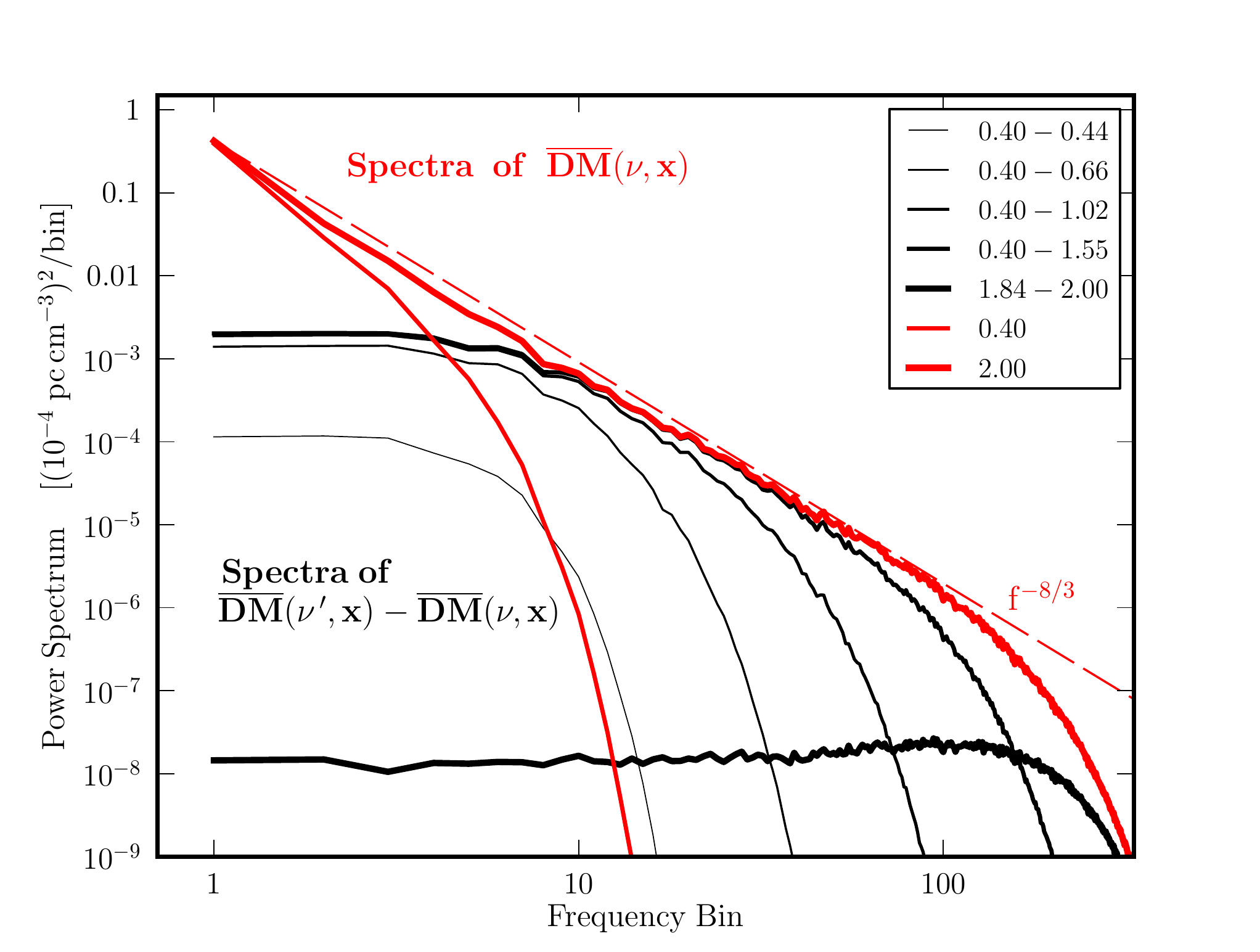}
\hspace{-0.4in}
\caption{\footnotesize  
Power spectra of dispersion measure variations $ \DMbar\ (\nu, \xvec)$ (red lines) and 
of dispersion-measure differences between frequencies $\DMbar(\nul) - \DMbar(\nuh)$ as designated
in the legend.  The results are based on 1000 realizations of a screen with $\phi_F = 30$~rad and a Kolmogorov
spectrum, which yields a power spectrum for DM variations with a $-8/3$ slope, as indicated by the dashed line.
\label{fig:DMspectra}
}
\end{center}
\end{figure}

 Figure~\ref{fig:DMtnu} shows realizations of $\DM(\nu, t)$ at eight frequencies (left-hand panel) obtained using  a phase screen for a Kolmogorov spectrum ($\beta = 11/3$).   The figure demonstrates that variations in \DM\ 
on long time scales  are similar over  a 5:1 frequency range,  but  differ on short time scales, with a more
rapid variation at higher frequencies.
  The variation in DM  has a very
`red'  power spectrum and is  a non-stationary process.   However, the DM differences between pairs of frequencies (right-hand panel) have  flatter spectra and show characteristic time scales associated with ray-path smoothing.   The autocorrelation widths are
roughly the geometric mean of the smoothing  lengths at each pair of frequencies.    These  smoothing lengths correspond to 
the times scales of refractive scintillations.  
Consequently the time series of DM differences  between a pair of frequencies is a  `red' process with a Gaussian-like power spectrum.    Figure~\ref{fig:DMspectra} shows power spectra of DM variations at two radio frequencies
(red curves) and power spectra of DM differences for several pairs of frequencies (black curves).


\subsection{ Frequency-Dependent  Averaging Over Ray Paths}
\label{sec:basics}

The measurable DM is an average over all ray paths that reach Earth.  At any location along the integration path,
the narrow ray-path bundle spans a transverse area described by $\xvec(z)$ that also depends on frequency. 
We define  a  frequency-dependent averaging function
$\Anuxz$ that is normalized to  unit area, $\int d\xvec\, \Anuxz = 1$. 
The averaged DM is the LOS integral of the convolution of the averaging function with the electron density,
\be
\DMbar(\nu, \xvec) = 
	\langle \DM(\xvec) \rangle 
	+  \int _0^D  d\zp\int_{-\infty}^{\infty} d\xvecp\, \Anuxxpz \delta n_e(\xvecp, \zp),
\label{eq:DMbar1}
\ee
where the frequency-independent term $\langle \DM(\xvec) \rangle$ is the ensemble average (denoted by
angular brackets) of the
integrated
electron density $\nebar$.   
An appropriate averaging function $\Anuxz$ is the scattered image of the pulsar that generally also includes refraction, which offsets the image from the actual pulsar direction. 
The scattered image $I(\thetavec)$ is the two-dimensional Fourier transform of the visibility function
$\Gamma(\bvec) = \exp(-D_\phi(\bvec)/2)$, where $D_\phi$ is the phase structure function  defined in 
Appendix~\ref{app:scatt}.
The phase structure function is 
the mean-square phase difference  of the wavefield  between  two points separated by distance $\bvec$
transverse to the LOS.  In accordance with our assumptions about the spectrum $\Pne$, $D_\phi$ depends
only on the magnitude of $\bvec$. 


 In the most
general case, the image can  be elliptical or split into multiple subimages.  
Elliptical images have been seen for some scattered sources while others are nearly axisymmetric.
An extreme case for a nearby
pulsar was given by \citet[][]{2010ApJ...708..232B}, who infer an image that is Gaussian-like but with 
$\sim 1$~\% of the flux spread into a long, linear image with high aspect ratio.   
Refraction angles are evidently smaller than scattering angles based on the lack of significant angular wandering
in VLBI images.  In addition, for media having wavenumber spectra like those that are consistent with a wide
range of scattering and scintillation measurements, refraction angles are expected to be small. 
Since our goal is to  describe the basic phenomenon of frequency-dependent DMs rather than address all possible varieties of scattering and refraction, the essential features are captured with a simplified approach that allows analytical tractability.  
The calculation undertaken here simply looks at the difference in electron column density that results from
the frequency dependence of the ray-path bundle that samples the medium.   The calculation of the 
arrival-time difference  does not take into account additional delays that result from lateral shifts of the ray-path
bundle due to refraction or changes  due to focusing and defocusing by quadratic phase changes.    
We emphasize therefore that our results  likely {\em underestimate} the total frequency-dependence of
DMs.

First we make some simple estimates of DM variations. 
From the  phase structure function $D_\phi(b)$ we show that the RMS  variation of $\DMbar$ across the 
screen-averaging area scales as
the {\it square} of the  RMS Fresnel phase $\phi_F$, and in the next section we derive a precise scaling law for the RMS difference in $\DMbar$ between two frequencies.     The DM structure function is equal to 
$D_\phi(b) / (\lambda r_e)^2$.  
{\it Without} any ray-path averaging, it can be written in terms of the RMS Fresnel phase as
\be
D_{\rm DM}(b) = \frac{\phi_F^2}{(\lambda r_e)^2} \left(\frac{b}{r_F} \right)^{\beta-2}. 
\ee
Averaging over ray-paths yields an RMS DM difference obtained
by averaging the DM structure
function over a circular area with radius $\sigma_X \sim r_F \phi_F^{2/(\beta-2)} \ge r_F$
(which can be derived using Eq.~\ref{eq:SMdef2}, \ref{eq:SMFres}, \ref{eq:sigtheta} and \ref{eq:sigX}). This prescription is applicable
only for moderate to strong scattering with $\phi_F \gtrsim 1$.   We discuss weak scattering in Section~\ref{sec:weak}. 
The averaging radius is essentially  the `refraction' scale for refractive scintillations, which is
equal to  the observed scattering angle
projected onto the screen at a distance $D^{\prime}$ from Earth (c.f. Figure~\ref{fig:raypaths});
it is  also the minimum scale
 for DM variations.  
This gives  for $\nu$ in GHz,  
\be
\DMbar_{\rm rms} \sim \phi_F^2 / \lambda r_e \sim \dDMone\ \nu \phi_F^2 .
\ee 
The scaling of $\DMbar_{\rm rms}$ as the square of   $\phi_F$ arises because  the screen phase and 
the averaging radius are both approximately linear in $\phi_F$.
The Fresnel phase scales with frequency as $\nu^{-17/12}$ so  $\DMbar_{\rm rms} \propto \nu^{-11/6}$.

In our detailed analysis, 
we use averaging functions that are {\em symmetric} and {\em concentric} at different frequencies.
In particular, we adopt a symmetric Gaussian function  for   $\Anuxz$, 
\be
\Anuxz = \left[2\pi \sigma_X^2(z,\nu)\right]^{-1} e^{-x^2/2\sigma_X^2(z,\nu)},
\label{eq:smooth}
\ee
where  the one-dimensional width  $\sigma_X$ is proportional to the observed scattering angle 
and therefore  scales with frequency as
$\sigma_X \propto \nu^{-\xtheta}$ with $\xtheta \approx 2$.   
We have confirmed numerically that the Gaussian averaging function yields nearly identical results to usage
of the scattering image appropriate for a Kolmogorov spectrum of fluctuations. 

\subsection{Two-frequency DM Differences}
\label{sec:2fddDM}

\begin{figure}[t!]
\begin{center}
\includegraphics[scale=0.5]
{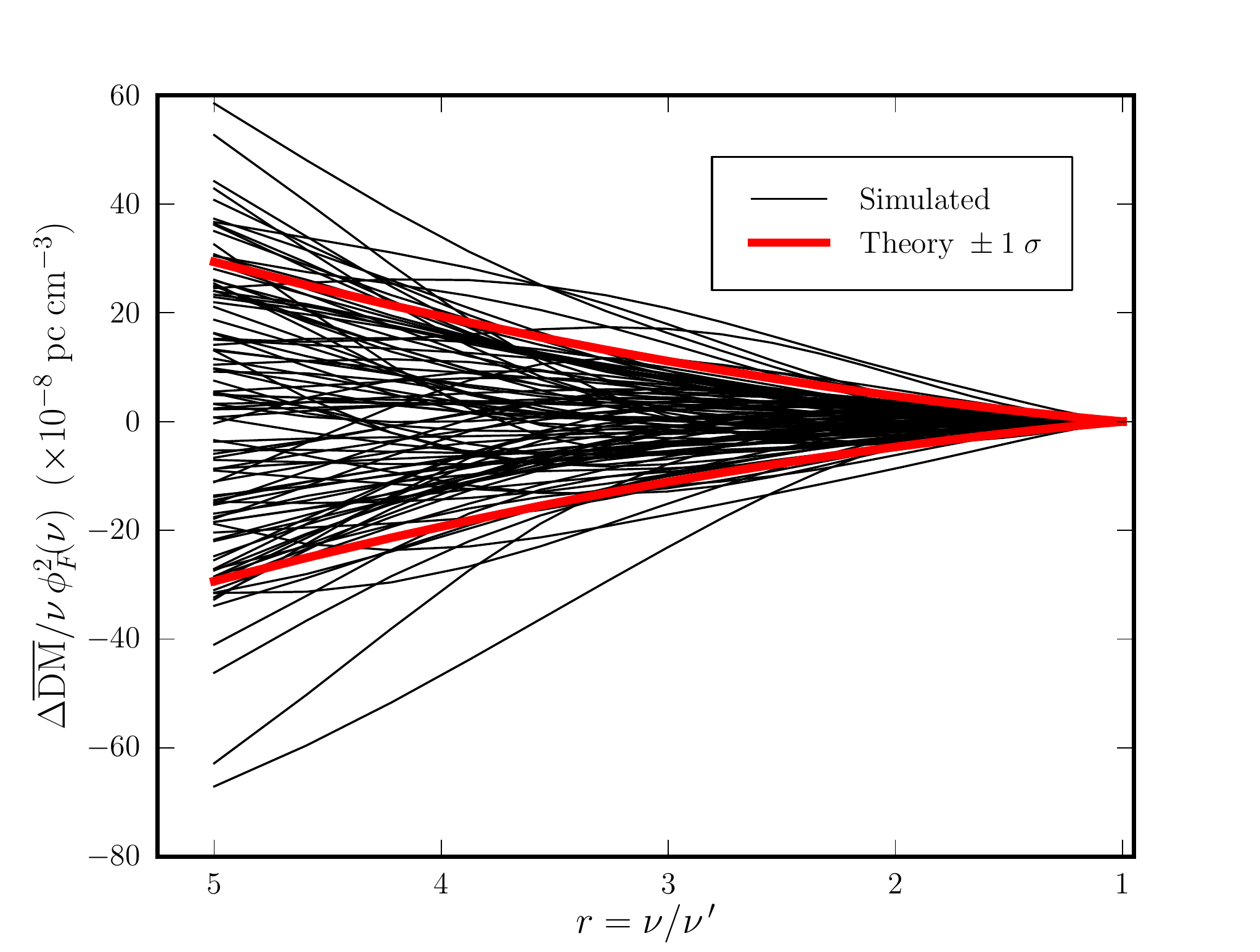}

\caption{\footnotesize Plot of $\Delta\DMbar$  normalized by $\nu\phi_F^2(\nu)$ vs frequency ratio $r = \nu/\nup$ for 100 realizations (black lines).
Also shown is $\pm F_\beta(r)$ scaled by the coefficient in Eq.~\ref{eq:rmsDMbarPhiF} (red lines), which is  the $\pm$~one standard deviation range of plotted values.   
\label{fig:ddDM}
}
\end{center}
\end{figure}


The  difference in $\DMbar$  between two frequencies $\nuh$ and $\nul$  (measured at the same location 
$\xvec$, corresponding to the same epoch), 
\be
\Delta \DMbar(\nuh, \nul, \xvec) = \DMbar(\nul, \xvec) - \DMbar(\nuh, \xvec),
\ee
has an RMS difference,
\be
\sigmadDMbar(\nuh, \nul)
	= \left\langle\left[\Delta \DMbar(\nuh, \nul, \,\xvec) \right]^2\right\rangle ^{1/2}.
\label{eq:ddm1a}
\label{eq:sf1}
 \ee
Electron-density wavenumber spectra   with $\beta > 2$ produce DM variations that are dominated by the largest scales.  For these $\beta$, the relevant scales range from 
the smoothing length $\sigma_X \approx \Dp {\theta_d}_\nu$   to the `outer scale'  of the spectrum that is likely determined by the sizes of structures and clouds in the ISM. The largest structures correspond to  time variations in DM  up to $10^4$ years or longer for parsec scales and  characteristic velocities 
$\sim 100$~km~s$^{-1}$.  We are concerned with much shorter time scales for which  the mean-square difference is an appropriate tool because large scale variations $\gg \sigma_X$ cancel out. 

In Appendix~\ref{app:2fDMSF} we derive the RMS DM difference using Gaussian smoothing functions and express the result in two forms, one that uses  the scattering measure
$SM = \int dz\ \cnsq(z)$  and a second that uses the Fresnel phase evaluated at the higher frequency,  
\be
\sigmadDMbar(\nuh, \nul)
	&=& F_\beta(\nuh/\nul) \times
	\left\{
		\begin{array}{ll}
			G_\beta Q_\beta \, r_e\ c^{\beta/2}  \, D^{(\beta-2)/2}   \nu^{-\beta/2}  \, \SM\,
				& \quad\quad\quad \text{Scattering Measure} 
				\label{eq:rmsDMbarSM}
				 \\
				 \\
			g_\beta \,q_\beta  \left(\frac{\displaystyle\nuh\phi_F^2(\nuh) }{\displaystyle c r_e}\right) 
				& \quad\quad\quad \text{Fresnel Phase}.
				\label{eq:rmsDMbarPhiF}
		\end{array}
	\right.	
	\label{eq:rmsDMbar}
\ee 
The dimensionless quantities $Q_\beta, q_\beta$ depend only on the wavenumber
spectrum while the dimensionless quantities $G_\beta, g_\beta$ depend on the LOS distribution of
$\cnsq$ (thin screen vs. statistically uniform medium, etc.) as well as on the spectrum. 
Values for  Kolmogorov media are given in Table~\ref{tab:parameters}.
All of the relative frequency dependence is contained in the function  $F_\beta(r)$, 
\be
F_\beta(r) = \left\{ 2^{(4-\beta)/2} \left[ 1 + r^{2\beta/(\beta-2)} \right]^{(\beta-2)/2} - r^{\beta} -1 \right\}^{1/2},
\label{eq:Fbeta}
\ee
where $r \equiv \nuh/\nul$. This vanishes for $r=1$, as expected and increases monotonically with $r$ for $2 < \beta < 4$. 

For a Kolmogorov spectrum,  

  \be
  \sigmadDMbar(\nuh, \nul)
	&=& F_{11/3}(r) \times
	\left\{
		\begin{array}{ll}
			\sigDMcoeffSM\,
			G_{11/3} \,  D^{5/6}   \nu^{-11/6}  \, \SM_{-3.5}\,
				& \quad\quad\quad \text{Scattering Measure} 
				\label{eq:rmsDMbarSMeval}
				\\
				\\
			\,
			\sigDMcoeffPhiF \,
			g_{11/3} \,  \displaystyle\left[\frac{\nuh\phi_F^2(\nuh) }{1000}\right]
				& \quad\quad\quad \text{Fresnel Phase}.
				\label{eq:rmsDMbarPhiFeval}
		\end{array}
	\right.	
\ee 
where SM is expressed in  units of $10^{-3.5}~\rm kpc~m^{-20/3}$, the distance
$D$ in kpc, and the frequency $\nuh$ in GHz; $\phi_F(\nuh)$ is in radians.    

The LOS is characterized by either \SM\ or $\phi_F$,  which can be estimated
from the  scintillation bandwidth $\dnud$ and  time scale $\Dtiss$. These are defined respectively  as 
the half-width at half-maximum of the intensity correlation function vs. frequency difference and as the half-width at $1/e$ of the intensity correlation function vs. time lag.\footnote{Usage of the half widths at half maximum and at $1/e$ for  the two cases is natural given the mathematical forms of the correlation functions for media with a square-law phase structure function.}
 For  a thin screen,  
\be
\SM_{-3.5} = 
	0.74\, \nu^{11/3} \left(\dnud D\right)^{-5/6} \left[ (\ds/D)(1-\ds/D) \right]^{-5/6} 
\ee
while for a statistically homogenous medium with  constant $\cnsq$, 
\be
\SM_{-3.5} = 2.26\, \nu^{11/3} \left(\dnud D  \right)^{-5/6}.
\ee
To get $\phi_F$ we use the  relationship  between  the pulse broadening time $\tau_d$ 
and  the scintillation bandwidth, 
$\tau_d = C_1/2\pi\dnud$, where $C_1$ is a constant of order unity \citep[][]{cr98, lr99}, 
\be
\phi_F(\nu) = \sqrt{2} 
		\left[\frac{C_1}{(D/D_s)(1-D_s/D)} \right]^{(\beta-2)/4}
		\left(\frac{\nu}{\dnud} \right)^{(\beta-2)/4}
		\approx 
		9.6~{\rm rad} 
		\left(\frac{\nu/\dnud}{100} \right)^{5/12},
\label{eq:phiF1}
\ee
where the quantity in square brackets $\sim {\cal O}(1)$ and the approximate expression is for a
Kolmogorov spectrum.   A ratio $\nu/\dnud \sim 100$ is of  order the value for nearby millisecond pulsars
at $\nu = 1$~GHz.  A similar expression can be written in terms of the diffractive scintillation time
scale $\Dtiss$ and the effective velocity $V_{\rm eff}$ by which the line of sight changes with time,
\be
\phi_F(\nu)  \approx 8~{\rm rad} \left(\sqrt{\nu} \,V_{\rm eff, 100} \Delta t_{\rm ISS, 1000} \right)^{-5/6},
\label{eq:phiF2}
\ee
with $V_{\rm eff} = 100V_{\rm eff, 100}$~km~s$^{-1}$, and $\Dtiss = 10^3 \Delta t_{\rm ISS, 1000}$~s.
A relation between scattering measure and $\phi_F$ is given in Eq.~\ref{eq:SMFres}.  

Figure~\ref{fig:ddDM} shows   $\Delta \DMbar$  
plotted against frequency ratio $r = \nuh/\nul$ (black lines) based on 
simulations of phase screens with $\phi_F = 30$~rad at 1 GHz.  
We also show  the predicted RMS  $\sigmadDMbar(\nuh, \nul)$ (red lines)  indicating statistical consistency between the theoretical and simulation results. Individual realizations show a tendency for $\Delta\DMbar$ to be dominated by
a linear dependence on  $r-1$, though there are counterexamples that show a significant quadratic dependence.
In Section~\ref{sec:timing} we utilize this approximate linearity as a basis for possible mitigation of the effect
in timing data.

\section{Implications for Timing Accuracy}
\label{sec:timing}

There are many complications to the estimation of arrival times. 
Recent work has addressed the removal of time-variable DMs 
\citep[e.g.][]{2014ApJ...790...93P, 2014MNRAS.441.2831L, 2014MNRAS.443.3752L} but no work to our 
knowledge has aimed at mitigating the frequency dependence.   
Other chromatic ISM effects include delays from diffraction and refraction that have been discussed elsewhere \citep[e.g.][]{fc90, r90, 2010arXiv1010.3785C}.  Also important is
the evolution with frequency of the intrinsic shape of a pulsar's pulse 
\citep[e.g.][]{1968Natur.220..676C, 1970PhDT.........8C, 2007MNRAS.377..677A, 2012A&A...543A..66H, 2014ApJ...790...93P, 2014MNRAS.443.3752L}, which produces a systematic
TOA error vs. frequency that is partially covariant with DM variations but is assumed to be independent of
epoch.      Since these and other effects are largely independent, their variances  add and therefore can be discussed separately from the frequency-dependent effect analyzed here.

We consider  multiple frequency measurements at a fixed epoch, so we drop any explicit
time dependence from $\DMbar$. A simple model for the TOA at frequency $\nu$ includes
dispersive delays and measurement errors as perturbations of the `true' TOA,  $t_\infty$, 
\be
t_\nu = \tinfty + \frac{\KDM \DMbar(\nu)}{\nu^2} + \epsilon_\nu.
\label{eq:tnu}
\ee
The quantity $\epsilon_\nu$ is an additive, frequency-dependent error that  is uncorrelated between
different frequencies for radiometer-noise but is highly correlated for intrinsic pulse jitter, at least over
modest frequency separations. 

\subsection{Dual Frequency Measurements}

An observing strategy  exploits the lever-arm provided by  widely-spaced frequencies  to obtain a high-precision estimate for DM, which then is used to estimate a DM-compensated arrival time, $\tinfty$.   Here we derive the resulting errors in DM  and $\tinfty$. 

Consider two  arrival times $t_\nuh, t_\nul$  measured at  frequencies $\nuh$ and $\nul < \nuh$ at the same epoch.  
The usual operational practice is to estimate  \DM\ (denoted by
the caret) by inverting Eq.~\ref{eq:tnu} under the assumption that \DM\ is frequency independent, 
\be
\widehat{\DM} = \frac{t_\nul - t_\nuh}{\KDM (\nul^{-2} - \nuh^{-2})}.	
\ee   
The dispersion delay is then removed from 
 the measured TOA $t_\nuh$ to estimate the infinite-frequency TOA, 
\be
\tinftyhat 
	= t_\nuh  - \frac{\KDM \widehat{\DM}}{\nuh^2} 
	= \tinfty + \frac{\KDM[\DMbar(\nuh) - \widehat{\DM}]}{\nuh^2} + \epsilon_\nuh.
\ee 
For a  frequency ratio $r = \nuh / \nul$, the difference between the estimated and true $\DMbar(\nuh)$ is
\be
\DMbar(\nuh) - \widehat{\DM} = 
		\left(\frac{r^2}{r^2 -1}\right) \left[\DMbar(\nuh) - \DMbar(\nul) \right]
		+
		\frac{\nu^2  \left(\epsilon_\nuh - \epsilon_\nul \right)}{\KDM}. 
\label{eq:dDMbar}
\ee
and the resulting  error in the infinite-frequency TOA is
\be
\delta \tinfty \equiv \tinfty - \tinftyhat = \frac{\KDM }{\nu^2} \left(\frac{r^2}{r^2-1} \right) \Delta\DMbar(\nuh, \nul)
	+ \left( \frac{\epsilon_\nul - r^2 \epsilon_\nuh }{r^2-1}\right).
\label{eq:TOAerror}
\ee
The estimator $\tinftyhat$  is  unbiased if the DM variations and the errors $\epsilon_\nu$ have zero mean values
over an ensemble.
The contribution to $\delta \tinfty$ from the measurement error $\epsilon_\nu$ at the higher frequency  is   enhanced by a factor $r^2$.   This means that the error at the lower frequency $\epsilon_\nul$ can be larger than the high-frequency error by a factor equal to a   modest fraction of $r^2$ and not unduly affect the precision
of $\tinftyhat$.

The combined RMS timing error $\sigma_{\tinftyhat}$ is  the quadratic sum of the individual RMS errors, 
\be
\sigma_{\tinftyhat} = \left(\sigma_{\tinfty, \delta\DMbar}^2 + \sigma_{\tinfty, \epsilon}^2\right)^{1/2}. 
\label{eq:sigtotal}
\ee

\subsubsection{TOA Error from  Frequency-dependent DMs}

The RMS DM difference  $\sigma_{\DMbar(\nuh, \nul)}$ defined in Eq.~\ref{eq:rmsDMbar}-\ref{eq:Fbeta}
implies an RMS error in TOA,  
\be
\sigma_{\tinfty, \delta\DMbar} =  \frac{r^2}{\vert r^2-1 \vert} \frac{\KDM\sigma_{\DMbar(\nuh, \nul)}}{\nuh^2} 
	&=&   E_\beta(r) \times
	\left\{
		\begin{array}{ll}
			\displaystyle
			(2\pi)^{-1} G_\beta Q_\beta \, r_e^2 \ c^{(\beta+2)/2}  \, D^{(\beta-2)/2}  \nu^{-(\beta+4)/2}  \, \SM,
				& \quad \text{Scattering Measure} 
				\label{eq:rmsTOA_SM1}
				 \\
				 \\
			g_\beta \,q_\beta  \left[\dfrac{\displaystyle\strut \phi_F^2(\nuh) }{\displaystyle \strut 2\pi \nuh }\right], 
				& \quad \text{Fresnel Phase}, 
				\label{eq:rmsTOA_PhiF1}
		\end{array}
	\right.
	\label{eq:sigtdm}	
\ee 
where the $r$-dependent factors have been combined into a timing-error function valid for  $2<\beta < 4$
(see Figure~\ref{fig:Ebeta}),
\be
E_\beta(r) = 
 		\dfrac{r^2 F_{\beta}(r)}{\vert r^2 - 1\vert}.
\ee

For a Kolmorovov spectrum, the RMS timing error is
  \be
  \sigma_{\tinfty, \delta\DMbar}
	&=& E_{11/3}(r) \times
	\left\{
		\begin{array}{ll}
			\sigTOAcoeffSM \,{\rm ns}\
			G_{11/3} \,  D^{5/6}   \nu^{-23/6}  \, \SM_{-3.5}\,
				& \quad\quad\quad \text{Scattering Measure} 
				\label{eq:rmsTOA_SMeval2}
				\\
				\\
			\sigTOAcoeffPhiF \, {\rm ns} \
			g_{11/3} \,q_{11/3}\, \left[ \displaystyle \dfrac{\nuh^{-1} \phi_F^2(\nuh) }{1000}\right]
				& \quad\quad\quad \text{Fresnel Phase}.
				\label{eq:rmsTOA_PhiFeval2}
		\end{array}
	\right.	
\ee 

\begin{figure}[t!]
\begin{center}
\includegraphics[scale=0.5]{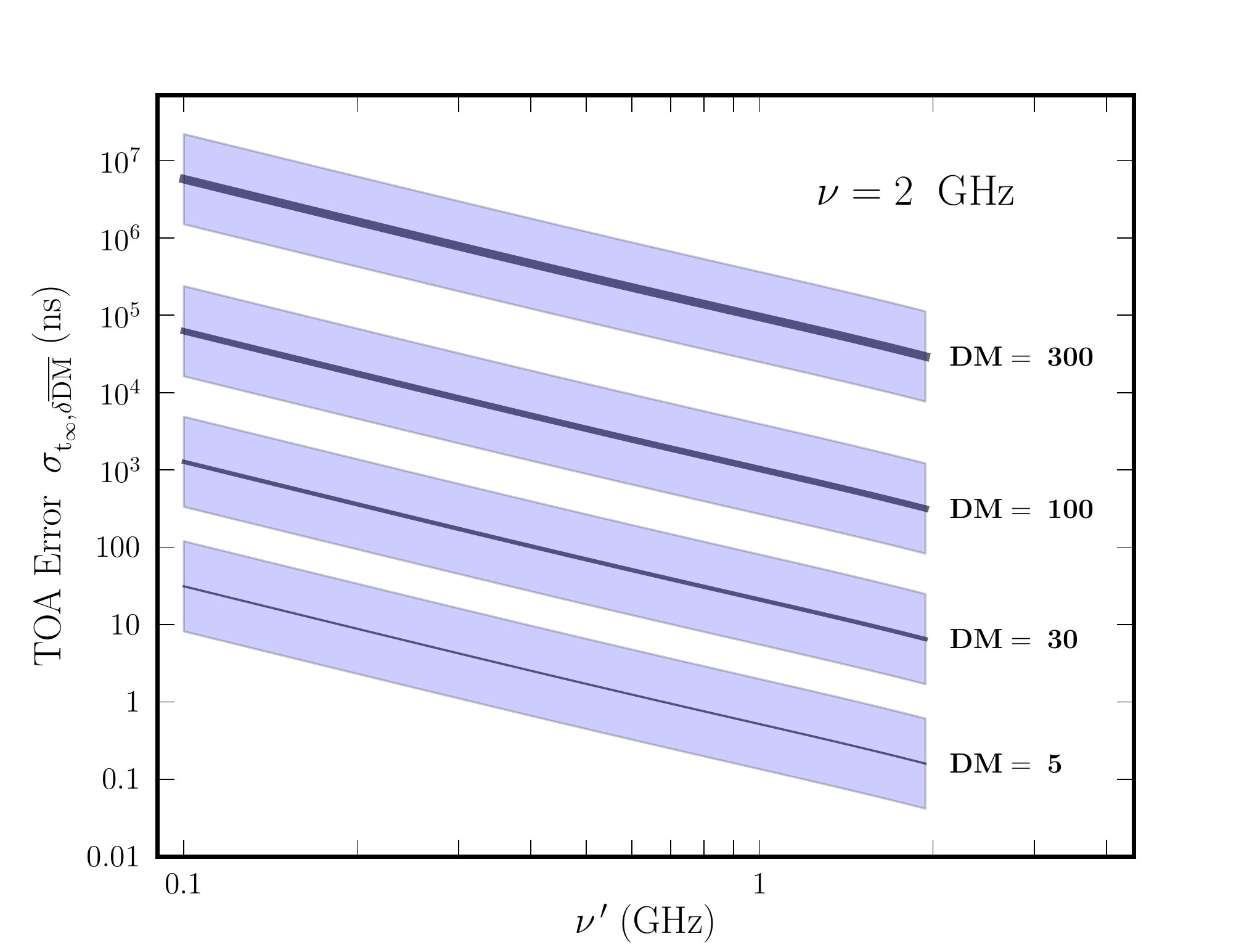}
\caption{\footnotesize  
RMS error in the DM-corrected TOA  $\tinfty$ vs. frequency for a fixed upper frequency of 2 GHz and  four values of
average DM, as labeled.  The curves for different DMs were calculated
using the pulse-broadening time $\tau_d$ from the scaling law in Eq.~7 of \citet[][]{2004ApJ...605..759B}
to obtain $\phi_F(\nuh)$ using Eq.~\ref{eq:phiF1} and then evaluating Eq.~\ref{eq:sigtdm} for different $r = \nuh/\nul$. 
The spread in values demarked by the shaded regions results from the  variation about the empirical mean relation
between  $\log_{10}\tau_d$  and $\log_{10}\DM$. 
While we show curves extending down to 0.1 GHz, pulse broadening will likely dominate the timing precision and will
render some pulsars undetectable at lower frequencies.
\label{fig:sigtoa_vs_frequency_variable_DMs}
}
\end{center}
\end{figure}

\begin{figure}[t!]
\begin{center}
\includegraphics[scale=0.4]{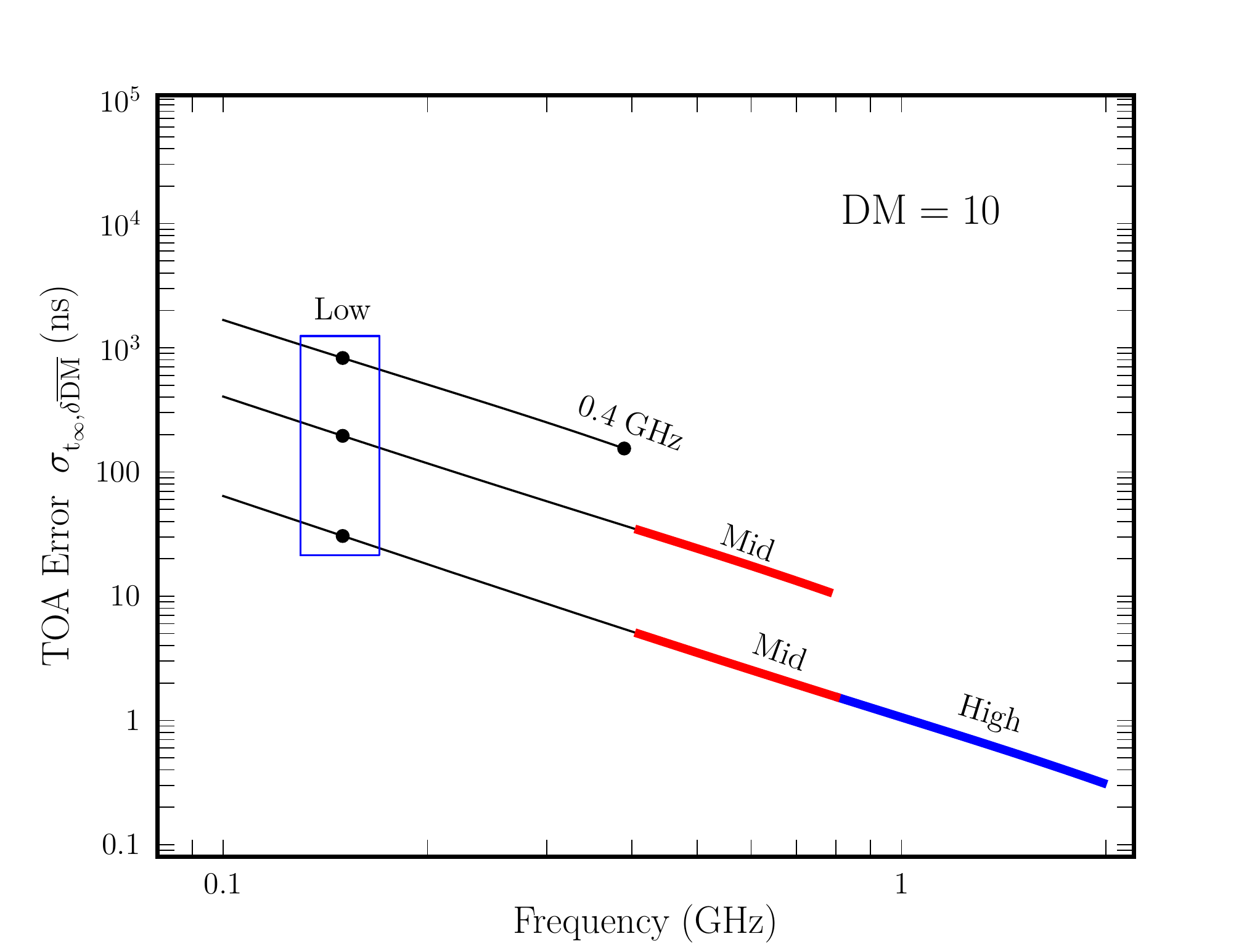}
\hspace {-0.3in}
\includegraphics[scale=0.4]{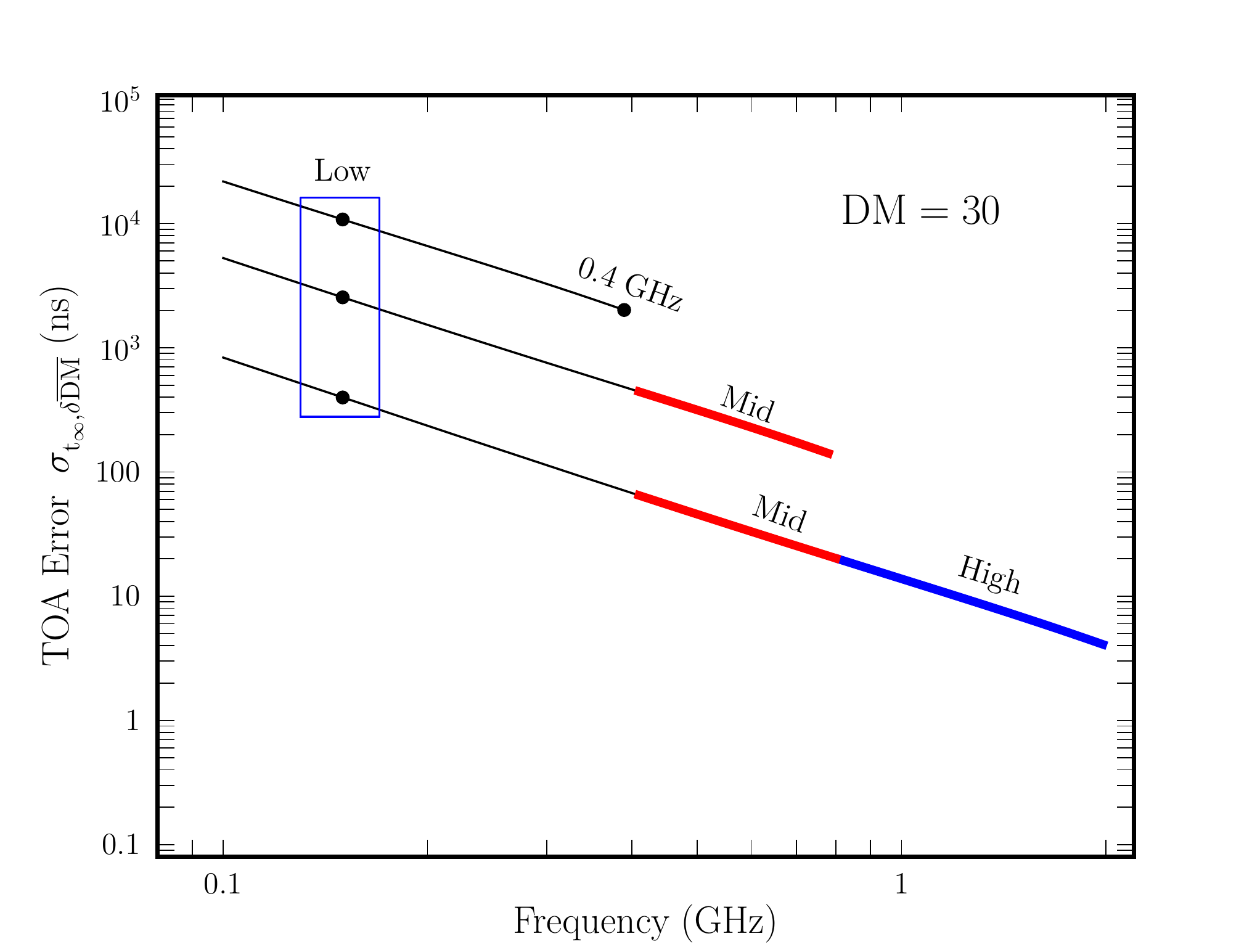}
\caption{\footnotesize
RMS $\tinfty$ vs. frequency $\nul$ for dual-frequency observations for pulsars with $DM = 10$~pc~cm$^{-3}$
(left) and $DM = 30$~pc~cm$^{-3}$ (right).   For each curve the rightmost frequency is the highest frequency
of the pair and the curve gives $\sigma_{\tinfty}(\nuh, \nul)$.  Lower $\nuh$ have higher $\sigma_{\tinfty}(\nuh, \nul)$.  Low, mid, and high-frequency bands are indicated.  
\label{fig:sigtoa_vs_telescopes}
}
\end{center}
\end{figure}

Figure~\ref{fig:sigtoa_vs_frequency_variable_DMs} shows $\sigma_{\tinfty, \delta\DMbar(\nul, \nuh)}$ vs $\nul$
for $\nuh = 2$~GHz for four values of average DM.  We calculated these curves by estimating the Fresnel
phase at the reference frequency $\nuh = 2$~GHz  using Eq.~\ref{eq:phiF1} and  then expressing the scintillation bandwidth at the highest frequency $\nu$ in terms of the pulse broadening
time using $\dnud = C_1 / 2\pi\taud$.  We get $\taud$ from  the empirical relation 
$\log_{10}\taud(\mu s) = -3.46 + 0.154x + 1.07x^2 - [2\beta/(\beta-2)] \log_{10}\nu$ for $\nu$ in GHz \citep[e.g.][]{2004ApJ...605..759B}.  The variation about this mean relation is $0.7$ in $\log_{10} \taud(\mu s)$. 
We use $\beta = 11/3$ though Bhat et al. find a best-fit value that is slightly greater\footnote{Specifically,
Bhat et al. fo
und a best fit  $\alpha = 3.86\pm 0.16$ for the coefficient of the $\log_{10}\nu$ term, which corresponds mathematically  to 
$\beta = 2\alpha/(\alpha-2 = 4.15\pm 0.17$, but this expression applies only for $\beta \le 4$. So a conservative interpretation
is that Bhat et al.'s result corresponds to $\beta > 11/3$.  Alternative interpretations  involve
other aspects of the scattering geometry rather than the form of the wavenumber spectrum.}.  The curves in the figure may therefore underestimate the timing error because we use $\nu = 2$~GHz. 
The figure demonstrates the strong dependence of the TOA error on chromatic DMs because $\taud$ and
thus $\phi_F$ increase nonlinearly with DM and the TOA error scales as the square of $\phi_F$.  

Figure~\ref{fig:sigtoa_vs_telescopes} complements Figure~\ref{fig:sigtoa_vs_frequency_variable_DMs} 
by showing  the TOA error for observations that involve different frequency ranges and for two different
but rather low values of DM (10 and 30~pc~cm$^{-3}$).   Each curve in the figure corresponds to two-frequency
observations where the higher frequency $\nuh$ is the right-most frequency (2, 0.8, and 0.4 GHz) and the lower
frequency $\nul$ is the horizontal axis.   Obtaining TOAs with $ < 0.1~\mu s$ precision  clearly requires a high frequency of no less than 0.8 GHz  for low-DM pulsars.   The curves presented here and in Figure~\ref{fig:sigtoa_vs_frequency_variable_DMs} demonstrate that low frequency telescopes such as LOFAR\footnote{http://www.astron.nl/radio-observatory/astronomers/technical-information/lofar-technical-information} need to be coupled with higher-frequency telescopes operating at $\gtrsim 2$~GHz to provide TOA errors $\lesssim 0.1~\mu s$ from chromatic DMs.  Note however, that the {\it total} DM-corrected
timing error also involves radiometer noise and pulse jitter, which favor wide frequency spans that are discussed next.




\subsubsection{Measurement Errors from Radiometer Noise, Pulse Jitter, and Diffractive Scintillations}

To treat any kind of measurement error, we  use a correlation function (normalized to unit maximum)   
$\rho_{\nuh\nul} = \langle \epsilon_\nuh \epsilon_{\nul} \rangle / \sigma_\nuh \sigma_\nul$ between the
errors at two frequencies.   
The RMS of the second term in Eq.~\ref{eq:TOAerror}    is then
\be
\sigma_{\tinfty, \epsilon} &=&  
	\frac{
		\left[ 
			r^4 \sigma_\nuh^2 + \sigma_\nul^2 - 2r^2\sigma_\nuh\sigma_\nul \rho_{\nuh\nul}
		\right]^{1/2}
		}
		{\vert r^2 - 1\vert}.
\ee


 

Radiometer noise ($\epsilon = $~rn) is uncorrelated between TOAs obtained using  non-overlapping
bandpasses, so $\rho_{\nuh\nul} = 0$ and
\be
\sigma_{\tinftyhat, \rm rn} = 
	\dfrac{\strut
			  \sqrt{ r^4 \sigma_\nuh^2 + \sigma_\nul^2 }
		}
		{\strut \vert r^2 - 1\vert}.
\label{eq:sigtrn}
\ee

In our analysis we consider wide frequency ranges with logarithmic spacings of frequencies and assume
that bandwidths are proportional to frequency.  This naturally leads to 
 a power-law scaling of the  TOA error from  radiometer noise, 
\be
\frac{\sigma_{\rm rn, \nup}}{\sigma_{\rm rn,\nu}} &=& \left(\frac{\nup}{\nu} \right)^{x_{\rm rn}} \equiv r^{-x_{\rm rn}}.
\ee
We then  rewrite Eq.~\ref{eq:sigtrn} as
\be
\sigma_{\tinftyhat, \rm rn} = 
	\dfrac
		{\sigma_{rn, \nu} \left(r^4 + r^{-x_{\rm rn}}\right)^{1/2} }
		{\strut \vert r^2 - 1\vert}.
\label{eq:sigtrn2}
\ee


Individual pulses show phase and amplitude jitter ($\epsilon =$~j) 
 that cause small shape changes  in the averages
of large numbers of pulses used to calculate TOAs.     The resulting TOA error  is correlated between frequencies,
sometimes highly so.  If perfectly correlated ($\rho_{\nuh\nul} = 1$) and identical,   jitter produces no error in 
$\widehat{\DM}$  because the TOAs move in tandem at the two frequencies  (c.f. Eq.~\ref{eq:dDMbar}).
The resulting TOA error from jitter is then simply $\sigma_j$.
However,  single pulses
and average profiles  evolve slowly with frequency, yielding random and systematic TOA errors, respectively. Generally the jitter correlation will be less than 100\%, yielding 
a larger contribution to the TOA error than from perfectly correlated jitter.   However, the frequency separation
over which single pulses decorrelate is large for the few cases that have been studied.  These 
include the Crab pulsar \citep[][]{1999ApJ...517..460S} which decorrelates over about 0.7~GHz and the millisecond pulsar J0437-4715, which decorrelates over a 2-GHz bandwidth
\citep[][]{2014MNRAS.443.1463S}.
   For the brightest pulsars,   the  TOA errors from jitter and radiometer noise
are comparable \citep[][]{2012ApJ...761...64S, 2014ApJ...794...21D, 2014MNRAS.443.1463S} so when
pulse jitter is not completely correlated,   mis-estimation of DM is comparable to that from radiometer noise.

We adopt a power-law scaling analogous to that for radiometer noise, 
\be
\frac{\sigma_{\rm j, \nup}}{\sigma_{\rm j,\nu}} &=& \left(\frac{\nup}{\nu} \right)^{x_{\rm j}} \equiv r^{-x_{\rm j}}.
\ee
The resulting expression for the jitter-induced timing error for perfect correlation between frequencies  ($\rho_{\nuh\nul} = 1$) is
\be
\sigma_{\tinftyhat, \rm j} = 
	\frac
		{ \sigma_{j,\nu} \vert r^2 -  r^{-x_{\rm j}}\vert }
		{\vert r^2 - 1\vert}.
\label{eq:sigtj}
\ee


Diffractive scintillations (DISS)  have  correlation times and  bandwidths that range, respectively, 
 from minutes to hours
and $\sim 100$~kHz to 100s of MHz at 1 GHz for currently timed millisecond pulsars (MSPs),
which tend to have low DMs $\lesssim 30$~pc~cm$^{-3}$.   
  DISS causes TOA errors because the associated pulse broadening function --- the
scattering impulse response of the ISM that is convolved with a pulsar's pulse --- is stochastic on 
the DISS correlation time scale.   The RMS TOA error at an individual frequency is much smaller than those
from radiometer noise and jitter for nearby MSPs but can be comparable for more distant ones (Lam~et.~al, in preparation).  

Scintillations of a low-DM MSP will yield a non-zero correlation $\rho_{\nuh\nul}$ for observations
made nearly simultaneously (e.g. within less than one hour) and with frequencies separated by no more than a correlation bandwidth. Most current timing observations are made with large-enough bandwidths or frequency separations and many observations are made non-simultaneously, in which case $\rho_{\nuh\nul}=0$.    

\subsubsection{Systematic Errors from Profile Evolution}

Profile evolution ($\epsilon = $~'pe') is known to introduce systematic errors in TOAs because the average profile changes smoothly with frequency.  They  can also be described using Eq.~\ref{eq:TOAerror}.   However, unlike the random errors in the previous section, profile evolution can be mitigated by exploiting its apparent epoch independence \citep[e.g.][]{2014ApJ...790...93P} and thus differs from the chromatic DM effect that varies with epoch. 

\subsubsection{Assessment of Two-Frequency Timing}

It is often assumed that wider frequency separations yield
 more precise dispersion measures and arrival times. 
We assess this approach by considering the   timing error from 
frequency-dependent DMs combined with  measurement errors.  The results indicate that there can be diminishing
if not worsening returns once the frequency ratio  becomes larger than about a factor of two and if 
no mitigation procedure is used.  We note that the same affliction arises from evolution of pulse shapes with
frequency, as mentioned earlier. 

\begin{figure}[t!]
\begin{center}
\includegraphics[scale=0.43]{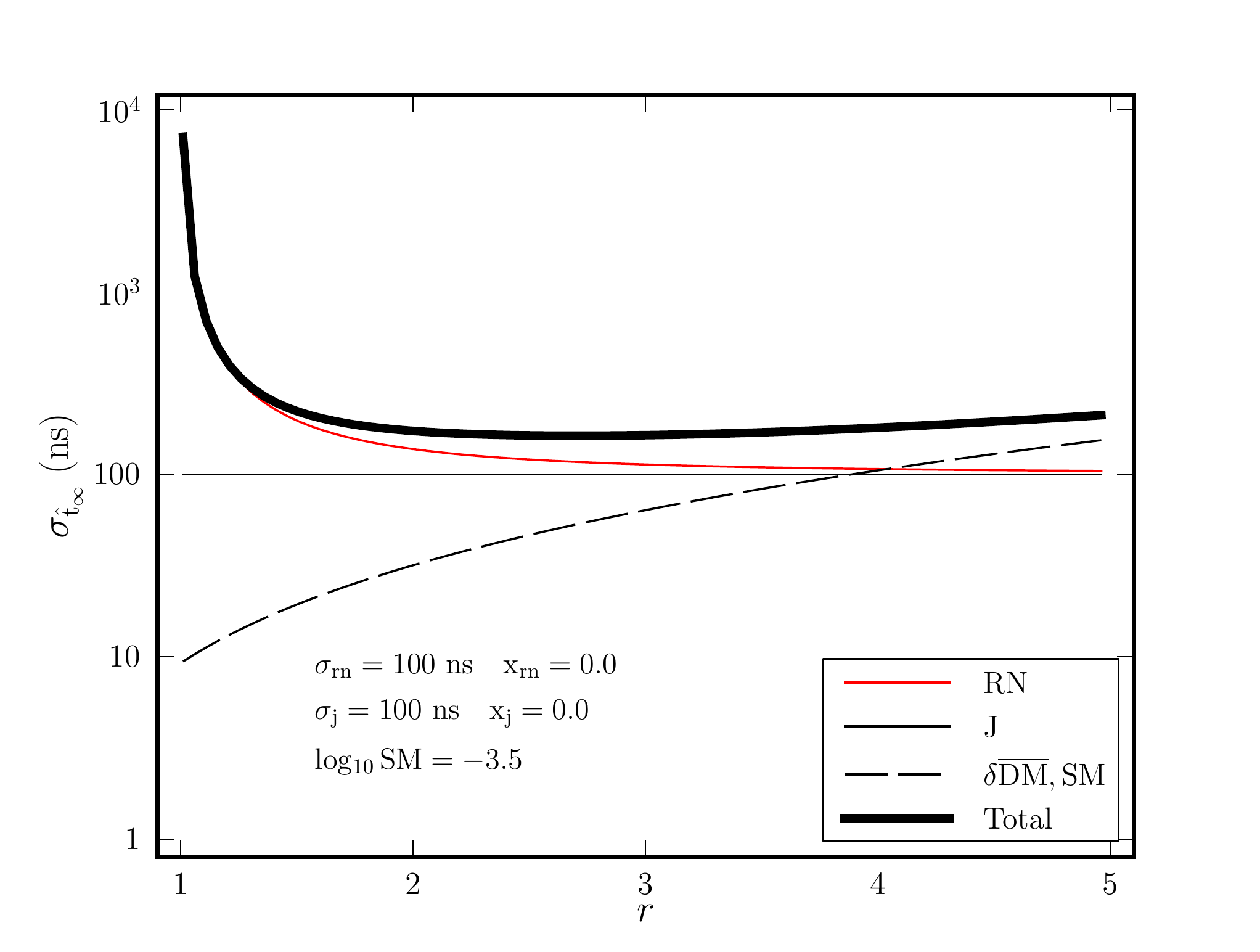}
\includegraphics[scale=0.43]{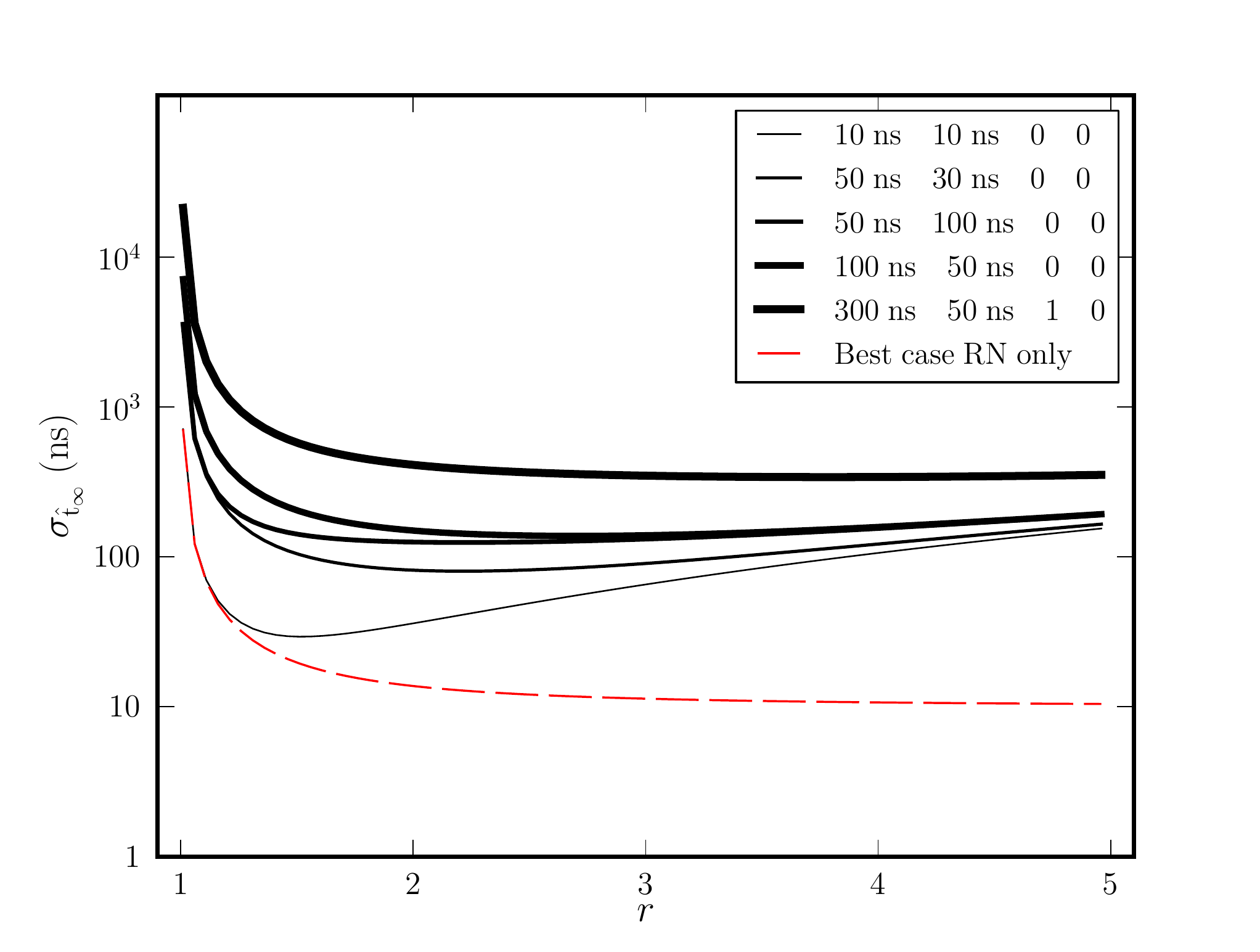}
\caption{\footnotesize  
(Left) RMS error of $\tinfty$ for two-frequency measurements as a function of
$r=\nul / \nuh$ for a fiducial (highest) frequency of 1~GHz.   Individual contributions are shown from radiometer noise, pulse jitter, and frequency-dependent DMs  along with the total (heavy black line).   
For this case we assume that the noise and jitter errors are both $100$~ns
and do not depend on frequency.   Results are not dramatically different for cases where they are 
frequency dependent. 
\label{fig:tinfty_breakdown}
(Right)
Total TOA error vs $r$ for six different mixtures of timing errors from radiometer noise, jitter, and
frequency dependent DMs.   The legend gives the TOA errors from  RMS noise $\sigma_{\rm t, rn}$, 
from RMS jitter, and their frequency scaling indices, $x_{\rm rn}$ and $x_{\rm j}$ respectively. The dashed
curve gives the TOA error if only radiometer noise with 10~ns RMS contributed to timing errors. 
\label{fig:tinfty6}
}
\end{center}
\end{figure}

In Figure~\ref{fig:tinfty_breakdown} we plot $\sigma_{\tinftyhat}$ from Eq.~\ref{eq:sigtotal} and the individual contributions to it
from frequency-dependent DMs, radiometer noise, and pulse jitter using Eq.~\ref{eq:rmsTOA_PhiFeval2}, \ref{eq:sigtrn},  and \ref{eq:sigtj}, respectively.   The specific case shown uses RMS values of 100~ns for both the noise and jitter contributions that are frequency independent ($x_{\rm rn} = x_{\rm j} = 0$).  The 
DM error was calculated for a scattering measure $\log_{10} \SM = -3.5$, a value that is typical of pulsars
within $1$~kpc of Earth.  The figure shows  basic features that are generic.  Rather than decreasing monotonically
or remaining flat, 
as expected from radiometer noise or pulse jitter alone, the TOA error reaches a minimum at $r\lesssim 2$
and then rises as the frequency-dependent DM contribution  begins to dominate.    Other cases
with different mixtures of radiometer noise and jitter and different indices $x_{\rm rn}$ and  $x_{\rm j} \ne 0$
are qualitatively similar.  

Other examples  are shown in Figure~\ref{fig:tinfty6}, including those with very small (10~ns) contributions
from noise and jitter.   While some cases --- those with large noise or jitter contributions --- do not show a minimum
in $r$, all  show much larger asymptotic TOA errors than expected from radiometer noise and jitter alone.
This implies that increases in bandwidth with wideband systems will provide diminishing returns unless the
frequency-dependent DM can be mitigated, as discussed below.

\begin{figure}[h!]
\begin{center}
\includegraphics[scale=0.41]
{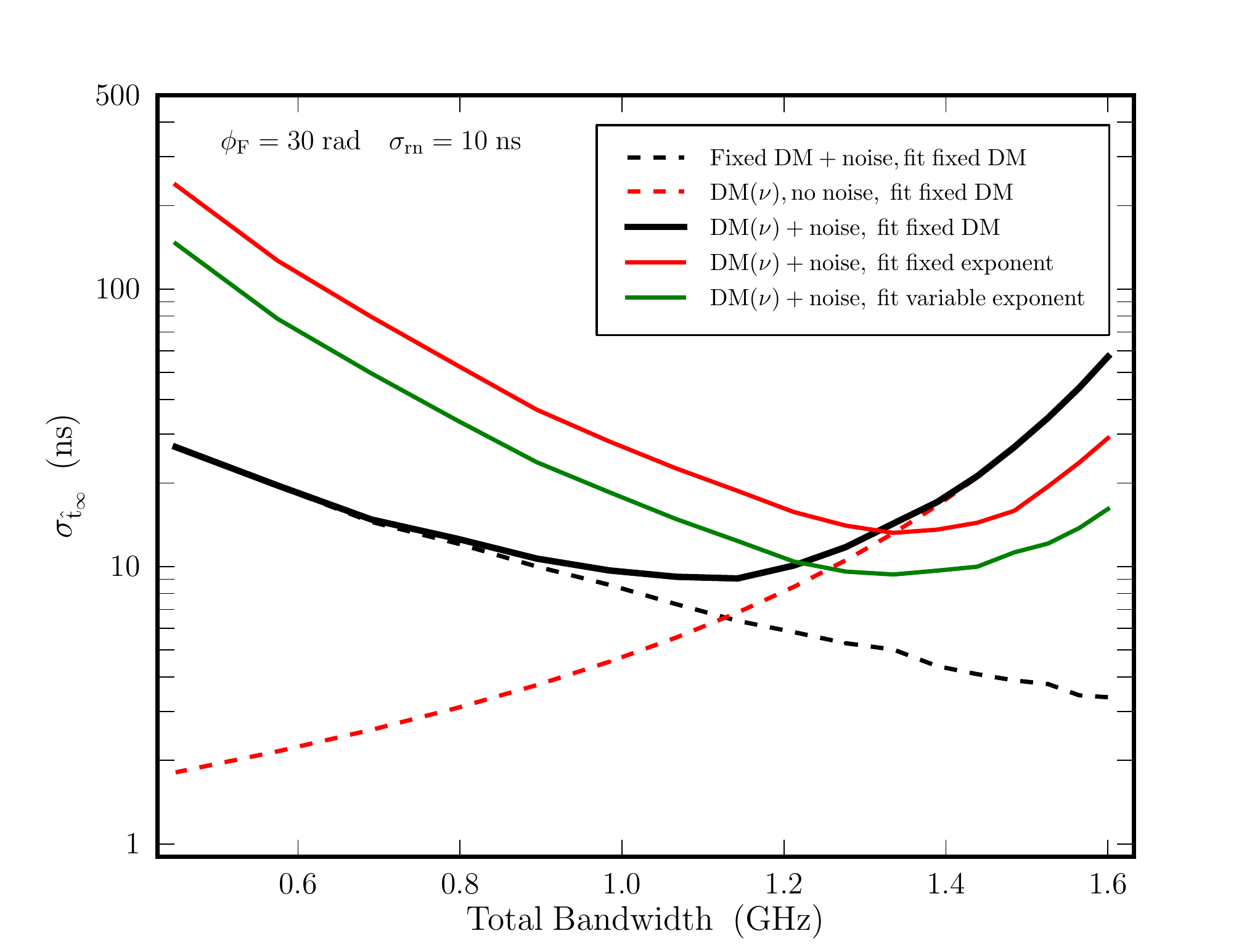}
\hspace{-0.3in}
\includegraphics[scale=0.41]
{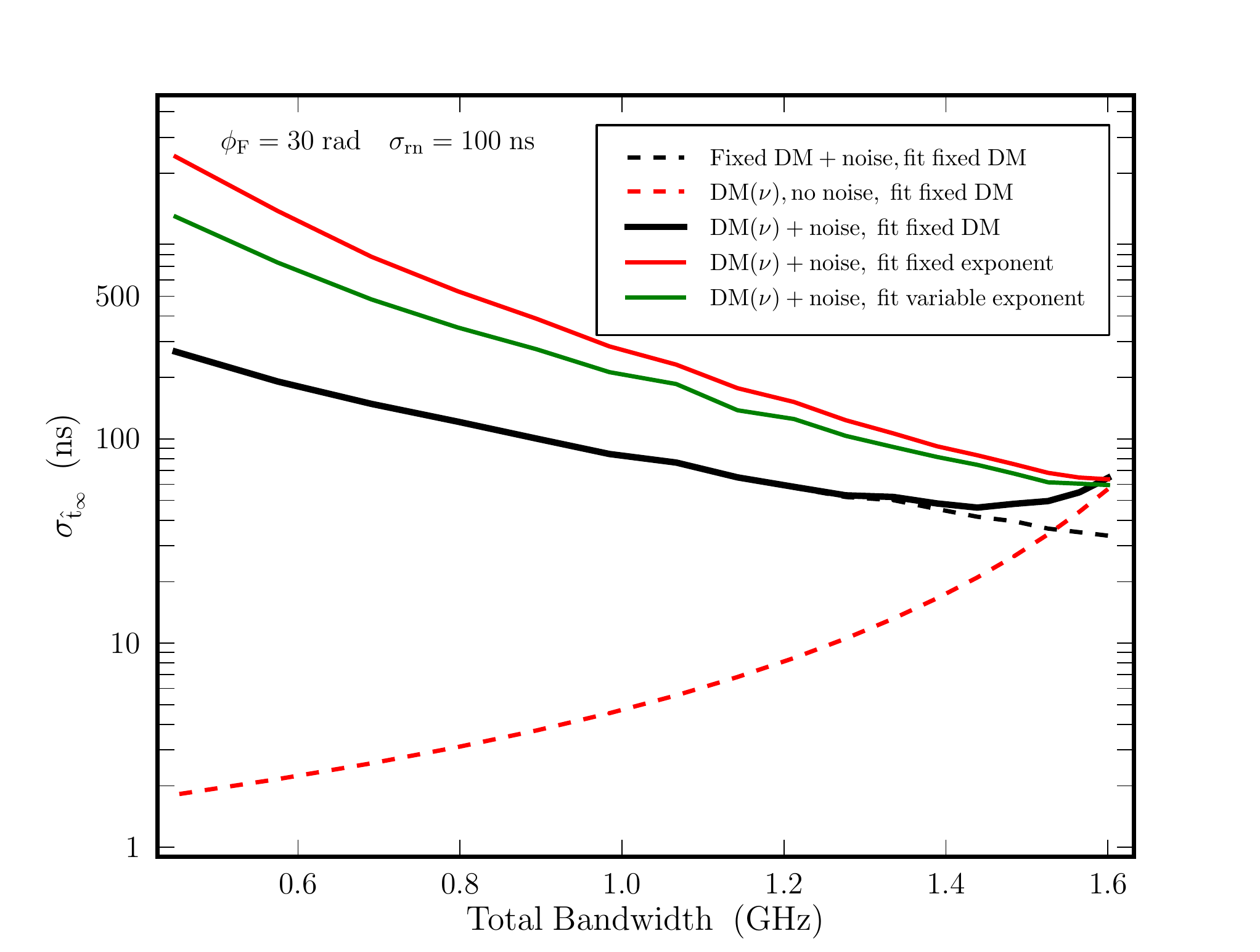}
\caption{\footnotesize  
RMS TOA error from wideband  fitting for DM and $\tinfty$  over a continuous frequency range from 0.4 to 2 GHz.
The phase screen has a Fresnel RMS phase of $30$~rad and the RMS radiometer noise is an optimistic 10 ns
(left) and 100~ns (right) that is assumed to be 
constant in frequency. A total of 2000 phase-screen realizations was used. 
The cases shown include    
(1)
Fixed true DM is fixed and a  timing model that includes only $\tinfty$ and  a $\nu^{-2}$ term; 
(2) Variable DM without noise added and a fit for fixed DM;
(3) Variable  DM with noise and a fit for fixed DM;
(4) As with (3) but with a fit that also includes a $\nu^{-X}$ term with fixed $X=3$;
and
(5) As with (4) but where the exponent $X$ is also fitted for. 
\label{fig:tinfty_wideband}
}
\end{center}
\end{figure}

\begin{figure}[h!]
\begin{center}
\includegraphics[scale=0.41]
{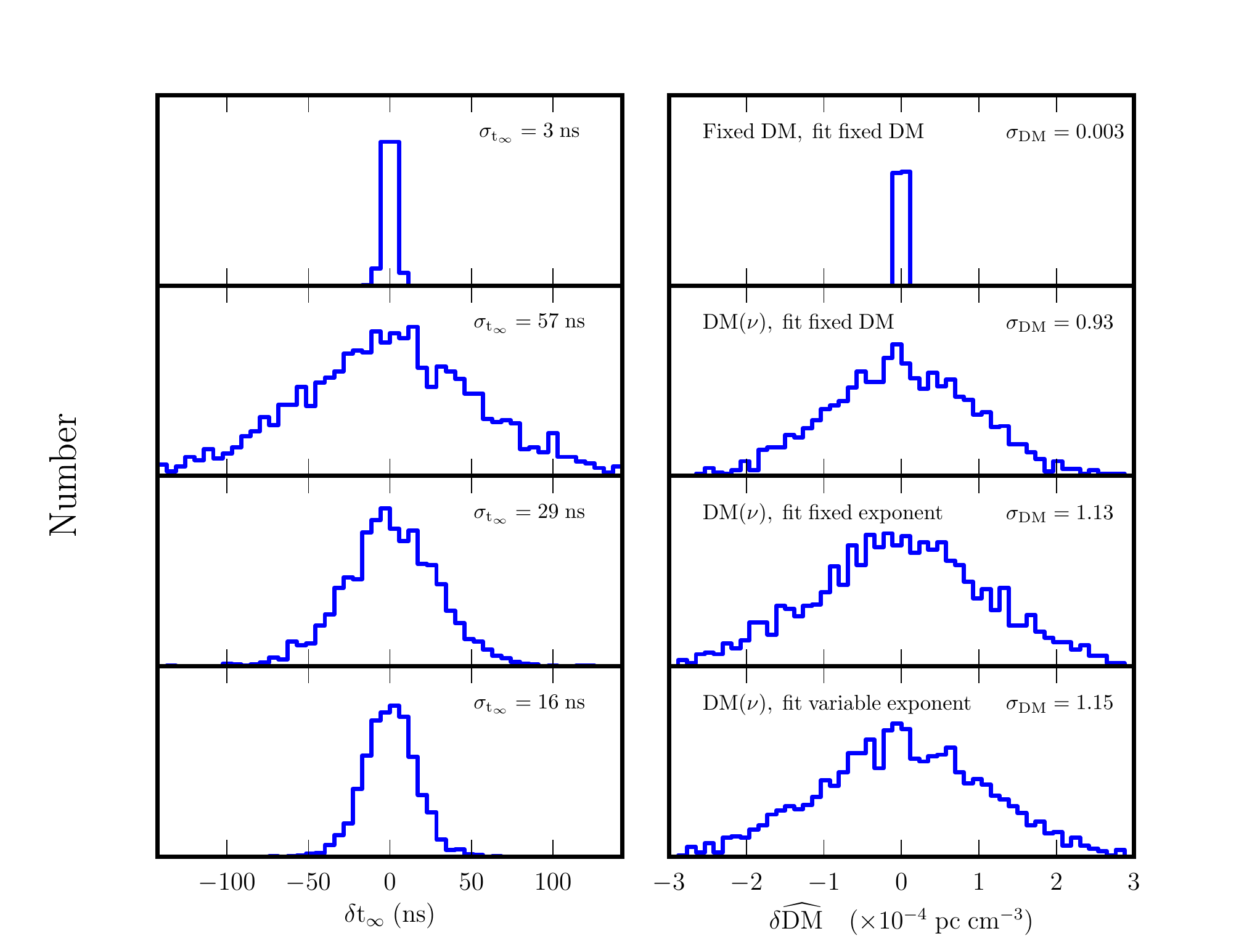}
\hspace{-0.3in}
\includegraphics[scale=0.41]
{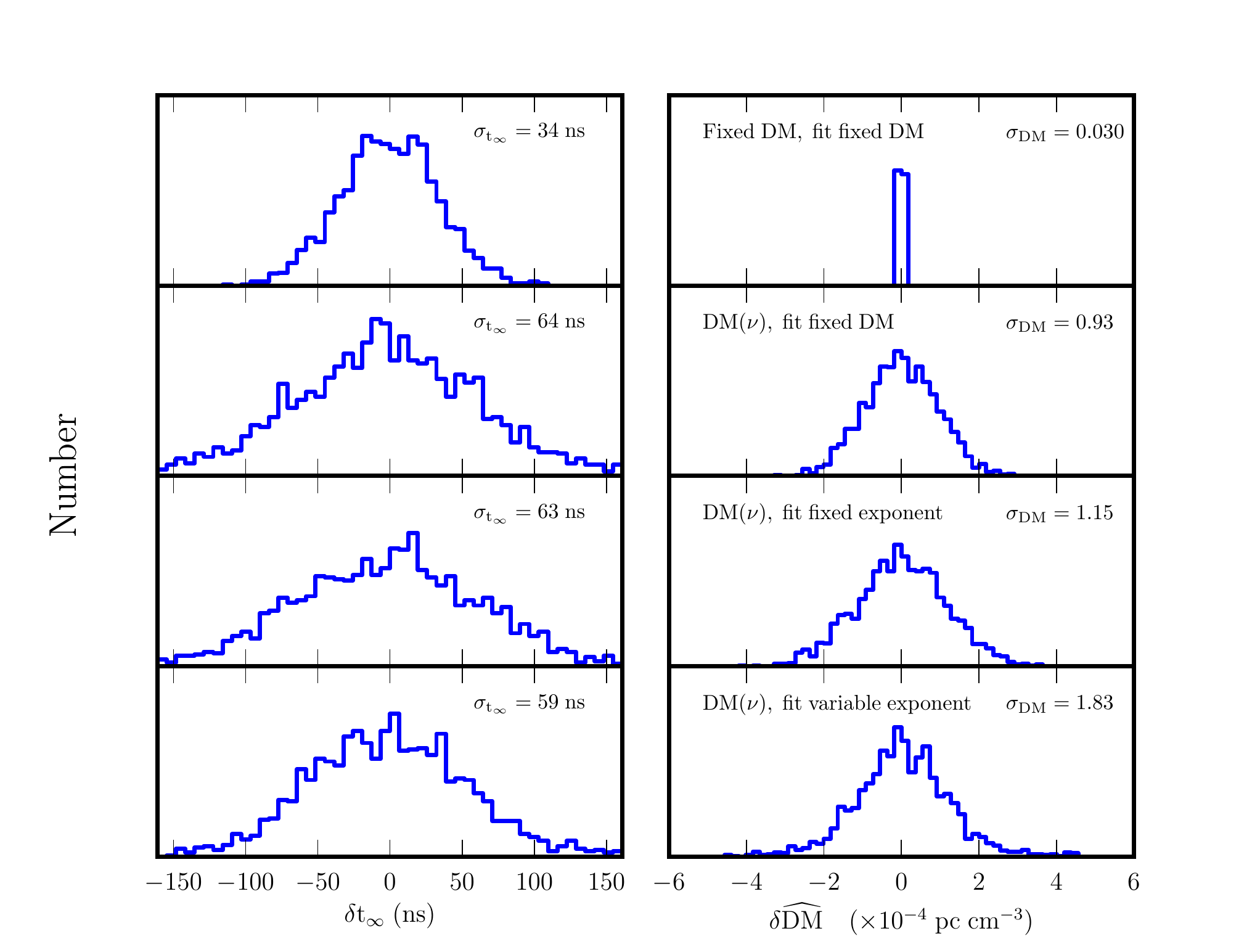}
\caption{\footnotesize  
Results from wideband  fitting for DM and $\tinfty$  over a continuous frequency range from 0.4 to 2 GHz.
The phase screen has a Fresnel RMS phase of $30$~rad and the RMS radiometer noise is 
constant in frequency: 10~ns (left set of panels) and 100~ns (right).
Left column: histograms of the error in $\tinfty$ based on 2000 realizations of a phase screen.
	RMS values for $\tinfty$ are given in each panel. 
Right column: histograms of the difference in DM from the value at the highest frequency.
	RMS values for $\delta\DMbar$ are given in units of $10^{-4}$~pc~cm$^{-3}$.   
First row: 
the true DM is fixed and the timing model includes only a $\nu^{-2}$ term; 
Second row:
same as the first case except that DM varies with frequency; 
Third row:
DM varies with frequency and the timing model includes 
	both a $\nu^{-2}$ and a $\nu^{-X}$ term with $X=3$ (fixed); 
and
Fourth row:
same as previous except that the exponent is also fitted for. 
\label{fig:tinfty_hists}
}
\end{center}
\end{figure}

\subsection{Wideband Timing Measurements}

Digital backend systems developed recently for pulsar observations have much larger frequency ranges
(1.8:1) than previous
systems and provide the opportunity to estimate TOAs over a continuous range of frequencies rather than 
using two narrowband frequencies with a wide separation \citep[e.g.][]{2014ApJ...790...93P}.      Even wider bandwidth systems are contemplated
with 4:1 (or larger) frequency ranges.  

We analyze wideband systems with arbitrary total bandwidths by  using standard   least-squares fitting methods
to fit data without and with frequency-dependent DMs. 
We let the  data vector $\Dvec$ comprise a set of TOAs $\{t_{\nu_k}, \, k=1, \ldots, N_\nu\}$  
and their measurement errors $\sigma_{\nu_k}$
for $N_\nu$  separate frequencies measured at the same epoch.  For simplicity,   we consider only radiometer noise in the wideband analysis, corresonding to a diagonal covariance matrix, 
 $\Carray =    {\rm diagonal} \left\{ \sigma_{\nu_k}^2 \right\}$. 
For a design matrix   $\Xarray$ and a linear  model $\Dvec = \Xarray \thetavec + \epsvec$, the solution
vector is 
$\thetavec =
\left(\Xarray^{\dagger}\Carray^{-1}\Xarray\right)^{-1}
\Xarray^{\dagger}\Carray^{-1}\Dvec$
and the covariance matrix for the parameters is 
${\bf P_{\thetavec}} = \left(\Xarray^{\dagger}\Carray^{-1}\Xarray\right)^{-1}$. 
 
With respect to the timing model of   Eq.~\ref{eq:tnu}, we consider four  alternatives:
\begin{enumerate}
\item
First is where the frequency-dependent DM is negligible so DM is constant in frequency and the
only errors are from radiometer noise.   This would be the case for a very low-DM pulsar or measurements at  high frequencies $\nu \gg 1$~GHz.  The data are fitted with a parameter vector 
$\thetavec = {\rm col} \left(t_{\infty}, \KDM \DM \right)$ and corresponding 
design matrix 
${\bf X} = {\rm matrix}(1 \,\,\, \nu_i^{-2}), i=1,\ldots,N_{\nu}$.  The solution vector is unbiased. 
\item
The second case is where the frequency dependence of DM is significant but the data are still fitted
with a constant DM model.    The results are generally biased. 
\item The third case is motivated by  recognizing in  Figure~\ref{fig:ddDM} that DM differences $\Delta \DMbar(\nul, \nuh)$
have a tendency to appear 
 roughly (but not exactly) linear in $r-1$; this suggests that a term in the fitting function
scaling as $\nu^{-3}$ 
will absorb much of the
effect and improve the estimate for $\tinfty$.  
\item The fourth case is the same as the third except the fitting function includes a $\nu^{-x}$  term
where $x$ is also fitted for  as a (nonlinear) parameter instead of fixing it at $x=3$. 
\end{enumerate}


Example results  in Figure~\ref{fig:tinfty_wideband}  show  the TOA error plotted against total bandwidth for  
$\nu = 2$~GHz (the highest frequency) with a total of 20 separate frequencies used.    DM variations are for a phase screen with 
$\phi_F = 30$~rad at $\nu = 1$~GHz. In the left-hand panel we have used an optimistic 10~ns for the radiometer
noise at the fiducial frequency of 1~GHz while in the right-hand panel we have used 100~ns. 
 We do not include
pulse jitter in these examples because results shown previously  indicate that it is secondary. 
Curves are shown for the four cases described above
 along with a fifth case that is an extreme version of case 2 where the noise is assumed negligible.
The results are similar to those obtained in the two-frequency case. Namely, for the 10-ns noise case,
increases in bandwidth
yield improvements until the bandwidth is about 1~GHz and then the results degrade if there is no
explicit fitting for the frequency-dependent DM.   With such fitting (cases 3 and 4 above), the bandwidth can
be increased another 20 to 40\% before the timing errors degrade.   For the larger 100-ns noise in the
right-hand panel,  TOA errors are dominated by radiometer noise except for the largest bandwidths for which
there is an increase in timing error.

Figure~\ref{fig:tinfty_hists} shows additional information on fitting results using histograms of the
errors in $\tinfty$ and $\DM$.  The two cases shown are for the same fits and data model used to produce
Figure~\ref{fig:tinfty_wideband}.   The top rows show fitting results when the DM is constant in frequency
and the sole source of fitting error is the additive noise.    Other rows show the results for the different methods
outlined above to deal with the frequency dependence of DM.  Overall,  Figures~\ref{fig:tinfty_wideband}
and \ref{fig:tinfty_hists} show that allowance for the frequency dependence can reduce the timing error but
cannot achieve the same results as for a constant DM.  
 
\subsection{Other Mitigation Methods}

So far we have discussed 
 estimation and removal of dispersion using data obtained  at a single epoch.  To date, most methods 
 used by groups aiming to detect gravitational waves 
 have used multiple epochs to estimate DM at the epoch of any particular arrival time.
 While using multiple epochs can be deleterious \citep[e.g.][]{2015ApJ...801..130L}, they can be implemented
 with algorithms that take into account the correlation times of  DM variations  that result from ray-path averaging. These times can be days to months or longer depending on the frequency and ISM along the line of sight.
 
\begin{figure}[h!]
\begin{center}
\includegraphics[scale=0.5]
{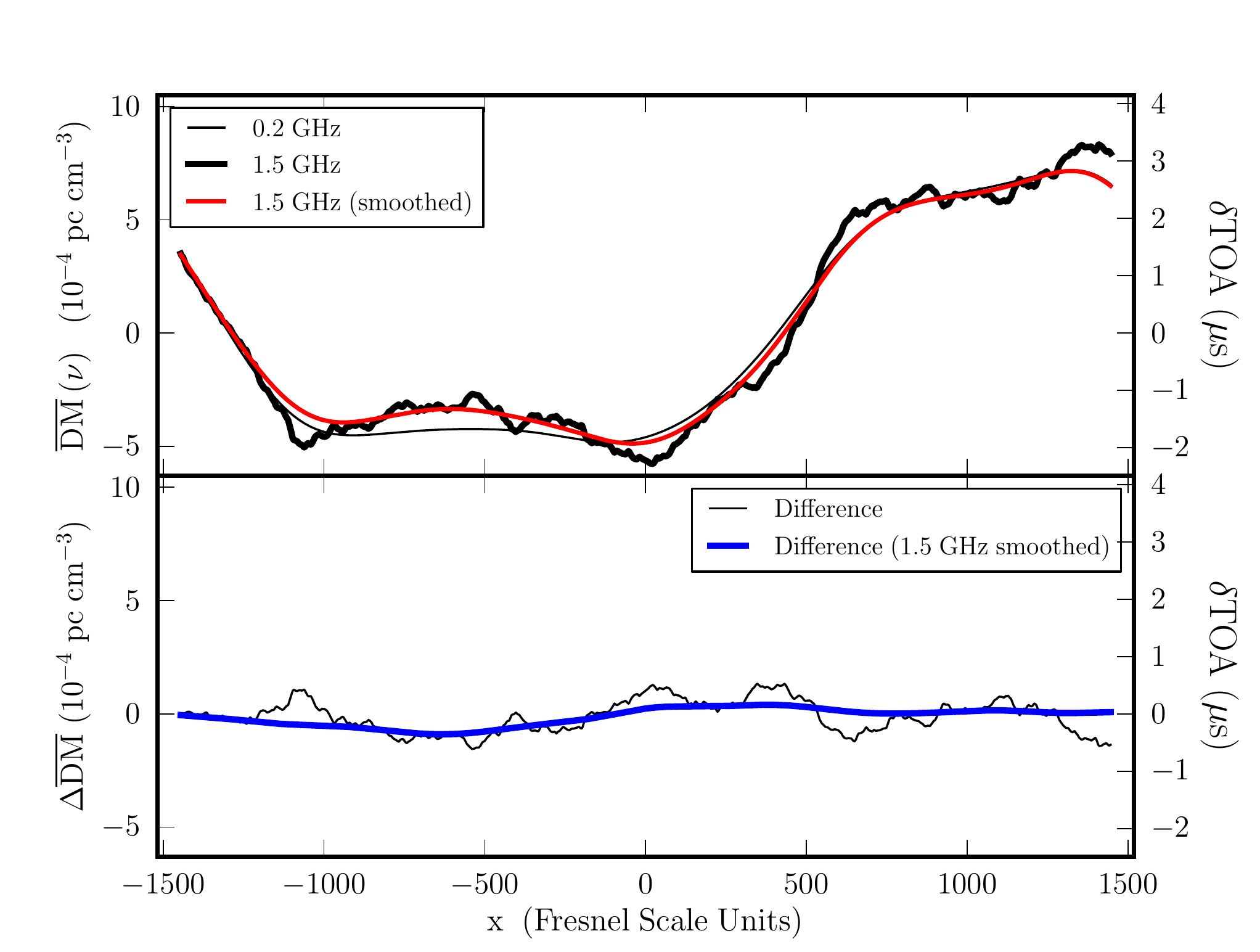}
\caption{\footnotesize 
Test of one-dimensional smoothing to estimate the two-dimensional averaging from scattering using
DM variations for a scattering screen with $\phi_F = 30$~rad.  The left-hand axis label is DM units and
the right-hand axis is in time-delay units.
(Top):  $\DMbar(\nu)$ vs. $x$ for two widely space frequencies, 0.2 and 1.5 GHz.
Also shown (red line) is the smoothed 1.5~GHz variation that best matches the 0.2 GHz curve using a
Gaussian smoothing function. 
(Bottom):  DM differences without and with smoothing of the high frequency variation.  
\label{fig:timeseries}
}
\end{center}
\end{figure}

 It is beyond the scope of this paper to develop a multi-epoch approach. However, we can illustrate
 a two-frequency  approach that smooths the DM time series at a high frequency  to match the characteristic
 time scale of the low-frequency DM variations.  Figure \ref{fig:timeseries},  shows DM variations at 0.2 and 1.5 GHz from a simulation for a phase screen with
$\phi_F = 30$~rad at 1~GHz.  In the top panel, the high-frequency time series has been optimally smoothed
using a Gaussian smoothing function that minimizes the mean-square difference with the low-frequency
time series.  The bottom panel of the figures shows the smoothed and unsmoothed DM difference between
the two frequencies.  While the differences have been reduced by smoothing the high-frequency DM,
they are not negligible because  one-dimensional smoothing  cannot
model the two-dimensional smoothing that occurs in the ISM from scattering.  Nonetheless, this approach
can be used as a mitigation procedure.

\subsection{Weak Scattering}
\label{sec:weak}

Our analysis applies to the strong scattering regime where $\phi_F > 1$.  Nearby pulsars observed at higher 
frequencies may be in weak scattering with $\phi_F < 1$.  In this regime, the scattered pulsar image consists
of the very compact image of the unscattered pulsar that contains a fraction $\sim 1-\phi_F^2$ of the total
flux combined with a scattered image with the remaining fraction.  For small $\phi_F$, the cross-sectional area
that is averaged is very small and the DM will be nearly achromatic.   This would be another advantage of
using high frequencies --- defined as those above the transition frequency to weak scattering.    The transition frequency is derived by requiring $\phi_F < 1$~rad and using  Eq.~\ref{eq:SMFres} to get
$\nu_{\rm trans} \approx 8~{\rm GHz}~\SM_{-3.5}^{6/17} D^{5/17}$.   For some pulsars, the transition 
frequency is $\lesssim 1$~GHz \citep[e.g.][]{2000ApJ...533..304R}. However, precise DM estimates require multiple-frequency (or wideband)
observations that almost always will use lower frequencies that are in the strong-scattering regime.   We therefore see no way to avoid the chromatic aspect of DMs in precision timing.

\section{Summary and Conclusions}
\label{sec:conclusions}

We have shown that dispersion measures are chromatic because microstructure in the interstellar electron density causes
multipath propagation that is strongly frequency dependent.  
We have characterized the effect in terms of the average over ray paths,  $\DMbar$, using an averaging kernel that 
is frequency dependent.  Results were given  for media having a power-law electron-density wavenumber spectrum
and for arbitrary variations in amplitude of the spectrum along the line of sight. 
We verified our analytical results with  simulations of phase screens with a Kolmogorov  spectrum.   
 Differences in $\DMbar$ between two frequencies scale as $\phi_F^2/\lambda r_e \propto \SM$,  where $\phi_F$ is
the RMS phase across a Fresnel scale at the highest frequency used and $\SM$ is the scattering measure.   
Observations at frequencies $\lesssim 1$~GHz will have $\phi_F \gg 1$ for most pulsars and $\phi_F \propto \nu^{-17/12}$.   
The specific variation of $\DMbar$ with frequency will remain constant in time over a refraction time scale of hours to months (or longer)  that is generally larger for larger-DM pulsars.  The joint variation of $\DMbar$ in time and frequency thus differs
from that of intrinsic profile evolution, which appears to be epoch independent for millisecond pulsars.   Longer period pulsars
show state switching (nulling, mode changes, etc.) on a wide range of time scales, but the switching statistics appear to
have stationary statistics.

As yet,  there has been no direct detection of  chromatic DMs.   This is not surprising because there has been no detailed analysis  prior to the present paper other than a brief discussion in \citet[][]{2010arXiv1010.3785C}, and its effects would be impossible to   identify uniquely in  dual-frequency TOA measurements typically employed. 
  However, departures from the
$\nu^{-2}$  dispersion delay appropriate for a tenuous, cold, unmagnetized plasma have been searched for
and used to  put limits on chromatic timing effects that scale differently than the standard
$\nu^{-2}$ dispersion law \citep[e.g.][and references therein]{2012A&A...543A..66H}.
 Such departures are complicated by profile evolution with frequency that is mitigated by identifying fiducial
pulse phases that yield consistency with the cold-plasma law.   Most of the objects analyzed this way
(pulsars and fast radio bursts of apparent extragalactic origin) have much coarser timing precision than the
MSPs used in pulsar timing arrays.    The best prospects for detection are from a bright, high-DM pulsar that has minimal profile evolution over the frequency range needed to probe DM variations.    Profile evolution may be disentangled from chromatic DMs by exploiting the lack of (or minimal) epoch dependence of profile evolution in comparison with variations in DM that have a characteristic correlation time for each frequency.    It is conceivable that some of the frequency-dependent timing variations observed from the MSP J1909$-$3744  \citep[e.g. Fig. 7 in ][]{2013PASA...30...17M} include the chromatic DM effect. 

Chromatic DMs have a significant impact on pulsar timing applications where sub-microsecond timing accuracy is needed,
such as detection of gravitational waves with pulsar-timing arrays and high-order general relativistic effects in binary pulsars.
All pulsar timing  applications depend on removing dispersion delays with high accuracy to estimate  TOAs  unaffected by propagation through the ISM, which we have called $\tinfty$.  We have analyzed  errors in $\tinfty$  resulting from  methodologies that assume DMs  are achromatic for  two cases, one  where TOAs are measured at pairs of widely separated frequencies and a second  that uses continuous, wideband systems.   Chromatic DMs introduce errors in $\tinfty$ that depend strongly on the pulsar's mean DM  and on the particular range of frequencies used.  For
nearby pulsars with $\DM \lesssim 30$~pc~cm$^{-3}$ and an octave frequency range with 
 1.5~GHz as the highest frequency, TOA errors solely from the chromatic part of DM are of order a few to hundreds of nanoseconds for observations extending down to 1~GHz or 0.2~GHz.    Timing errors increase rapidly with increasing  mean DM and timing residuals will be  correlated on time scales related to those of refractive interstellar scintillations. 
 
Timing errors   from radiometer noise and pulse jitter combined with chromatic DMs show a broad minimum as a function of total frequency range (or bandwidth) that is of order an octave in frequency.   This arises
because TOAs improve monotonically with bandwidth as far as radiometer noise is concerned, but the opposite
is true for  chromatic DMs.   The simplest prescription for optimizing TOA precision is to use an upper  frequency  that is as high as possible.      The choice of upper frequency is strongly  pulsar and telescope dependent.  




Chromatic DMs also need to be considered in combination with other chromatic effects, including
 intrinsic pulse profile evolution with frequency and additional interstellar
delays that result directly from scattering and refraction  that have different frequency dependences than any of the
DM effects.   A comprehensive assessment of effects like that in \citet[][]{2010arXiv1010.3785C} 
 that includes the chromatic DM effect is deferred to a separate paper.
 Even though the timing errors from chromatic DMs are smaller than other effects, they nonetheless may
inhibit improvements in timing accuracy that  otherwise might be obtainable.  It is therefore important to confirm 
that chromatic DMs are  present in timing data at predicted levels and develop ways to mitigate them, if possible.  If not, they need to be part of the noise model for timing analysis and incorporated into the covariance matrix used in model fitting.

In a separate article we will assess the role of chromatic DMs for all objects currently being observed in pulsar timing array
campaigns to detect gravitational waves and we will also assess different methodologies for using existing and future telescopes.  We will also identify particular pulsars that are good candidates for direct detection
of DM chromaticity.


\acknowledgements
We thank colleagues in the NANOGrav collaboration, in particular the Interstellar Mitigation Working Group,
and colleagues in the European Pulsar Timing Array and Parkes Pulsar Timing Array collaborations
 for useful discussions that stimulated the work reported here.
JMC's work on pulsar timing at Cornell University is supported in part by NSF PIRE program award number 0968296.
RMS acknowledges travel support through a John Philip early career
research award from the CSIRO.
DRS gratefully acknowledges research support from NSF grant 1313120.

\begin{appendix}
\section{Phase Structure Function and Scattering Angle}
\label{app:scatt}


The phase structure function  is the integral from a point source to an observer at distance $D$ from a point source, 
\be
D_\phi(\bvec) =  \left\langle \left[\phi(\xvec) - \phi(\xvec + \bvec)\right]^2 \right\rangle =  4\pi (r_e \lambda)^2 \int_0^D dz\, \int d\qperpvec\, \Pne(\qperpvec,  z) \left(1 - e^{i \displaystyle \qperpvec\cdot\bvec z / D}\right)
\ee
where $\bvec$ is a spatial offset and $\qperpvec$ a vector  wavenumber transverse to the line of sight.  
This form applies to a wavenumber spectrum 
$\Pne$ whose    extent in wavenumber is much narrower than $D^{-1}$,  has a shape independent of $z$,  and an amplitude that varies slowly with $z$. 
 Normalization is  so that the mean-square electron density is the integral of
 $\Pne$ over all wavenumbers, and $\int_{-\infty}^{\infty} \!dz\, e^{iq_z z} = 2\pi\delta(q_z)$. 
 
Scattering measurements indicate various degrees of anisotropy of density fluctuations,  but the isotropic case is easier to analyze.  To treat the anisotropic case is  tedious and does
not add any further insights to the results obtained for the isotropic case. 
For isotropic irregularities only  the magnitudes of  $\bvec$ and  $\qperpvec$ matter, yielding
\be
D_\phi(b) = 8\pi^2 (r_e \lambda)^2 \int_0^D dz\, \int dq_\perp \, q_\perp \Pne(q_\perp, z) [1 - J_0(q_\perp bz/D)],
\label{eq:Dphi1}
\ee
where $J_0$ is the Bessel function of the first kind. 
We adopt a power-law wavenumber spectrum, 
\be
\Pne(\qperp, z) = \cnsq(z) q^{-\beta}, \quad\quad q_0 \le q \le q_1,
\label{eq:pne2}
\ee
where $\cnsq(z)$ varies slowly with $z$ on length scales much larger than the outer scale, $2\pi/q_0$. 
For $b \ll q_1^{-1}$  (i.e. $b$ smaller than the inner scale),  the phase structure function is quadratic in $b$
while for $b \gg q_0^{-1}$ it asymptotes to twice the total variance of the phase.
In the intermediate regime where 
 $q_1^{-1} \ll b \ll q_0^{-1}$, 
\be
D_\phi(b) \approx  f_\beta (\lambda r_e)^2 \SMeff\ b^{\beta-2},
\label{eq:Dphi}
\ee
 where (Cordes \& Rickett 1998; Eq. B6)
\be
f_\beta = \frac{8\pi^2}{(\beta-2) 2^{\beta-2}}
\frac{\Gamma(2-\beta/2)}{\Gamma(\beta/2)},
\ee
where the  effective scattering measure  is  the line-of-sight weighted integral of $\cnsq$,
\be
\SMeff = \int_0^D dz\, \cnsq(z) \left(\frac{z}{D}\right)^{\beta -2}
	\equiv \SM \times \frac{\int dz\, \cnsq(z) (z/D)^{\beta-2}} {\int dz\, \cnsq(z)}
	\equiv \SM \left\langle (z/D)^{\beta-2}\right\rangle_{\cnsq},
\label{eq:SMeff}
\ee
where the angular brackets denote an average over the LOS using  $\cnsq(z)$ as 
a weighting function and  the scattering measure is
\be
\SM \equiv \int_0^D dz\, \cnsq(z).
\label{eq:SMdef}
\ee
For a screen, $\SMeff = (\ds/D)^{\beta-2}\ \SM$ and for a uniform medium with $\cnsq = $~constant,
$\SMeff = \SM/(\beta-1)$.  A plane wave incident on a foreground scattering medium so that $\ds/D \to 1$
gives $\SMeff\equiv SM$ (e.g. for an extragalactic pulse incident on the Milky Way). 
Values for $\SMeff/SM$ are given in Table~\ref{tab:parameters} for a Kolmogorov spectrum  along with other parameters quantities.

Alternatively we can express the phase structure function in terms of the RMS phase $\phi_F$ across 
a Fresnel radius $r_F$ in the screen.  We define the Fresnel scale using $r_F^2 = [(\lambda D)/2\pi] \Deff/D$ where for a thin screen $\Deff = \ds \Dp / D = D(\ds/D)(1-\ds/D)$.  For a statistically uniform medium we assume
$\Deff = D/4$.  Taking into account  that a transverse scale $b$ at an observer's position corresponds
to a scale $zb/D$ at a position $z$ along the LOS, we have for a thin screen at $z=\ds$, 
\be
D_\phi(b) = \phi_F^2 
	\left(\frac{\ds}{ D}\right)^{\beta-2}
	\left(\frac{b}{ r_F}\right)^{\beta-2}.
\label{eq:SMdef2}
\ee
From this and Eq.~\ref{eq:SMdef} we  solve for $\SM$ in terms of $\phi_F$,  
\be
\SM = 	\frac{1}{f_\beta r_F^{\beta-2} } 
		\frac{\phi_F^2}{(\lambda r_e)^2}. 
\label{eq:SMFres}
\ee
 In the following we assume the same relation holds generally though we have derived it from the thin-screen case.

The scattered image of a point source has longer tails  than a Gaussian function for $\beta < 4$ and an inner scale $2\pi/q_1$  much smaller than the Fresnel scale \citep[e.g.][]{r90}.  However it is convenient to characterize the main part of the image with an equivalent Gaussian  whose visibility function has the same $1/e$ width.  
This defines the spatial scale $b_e$ using $D_\phi(b_e) = 2$,
\be
b_e = 
	\left[ 
		\frac{2}{f_\beta (\lambda r_e)^2 \SMeff}
	\right]^{1/(\beta-2)},
\ee
from which the RMS angular size $\sigtheta$, the $1/e$ half width, and the full width at half maximum (FWHM) are
calculated as
\be
\sigtheta = \frac{\theta_e}{\sqrt{2}} = \frac{\FWHM}{2\sqrt{2\ln 2} } =  \frac{\lambda}{\sqrt{2}\pi b_e}. 
\ee
The power-law spectrum yields an RMS angular size 
that we factor into  a scattering size $\sigthetas$ and a geometry-dependent factor, 
\be
\sigtheta  \equiv \sigthetas \left(\frac{\SMeff}{\SM}\right)^{1/(\beta-2)}, 
	\quad\quad\quad 
	\sigthetas = 
	\frac{1}{\pi}
	\left[
		\frac{\lambda^{\beta} f_\beta r_e^2 \SM}{2^{\beta/2}}
	\right]^{1/(\beta-2)}.
	\label{eq:sigtheta}
\ee

\section{Two-frequency Cross Correlation and Structure Function}
\label{app:2fDMSF}

\begin{figure}[t!]
\begin{center}
\includegraphics[scale=0.40]{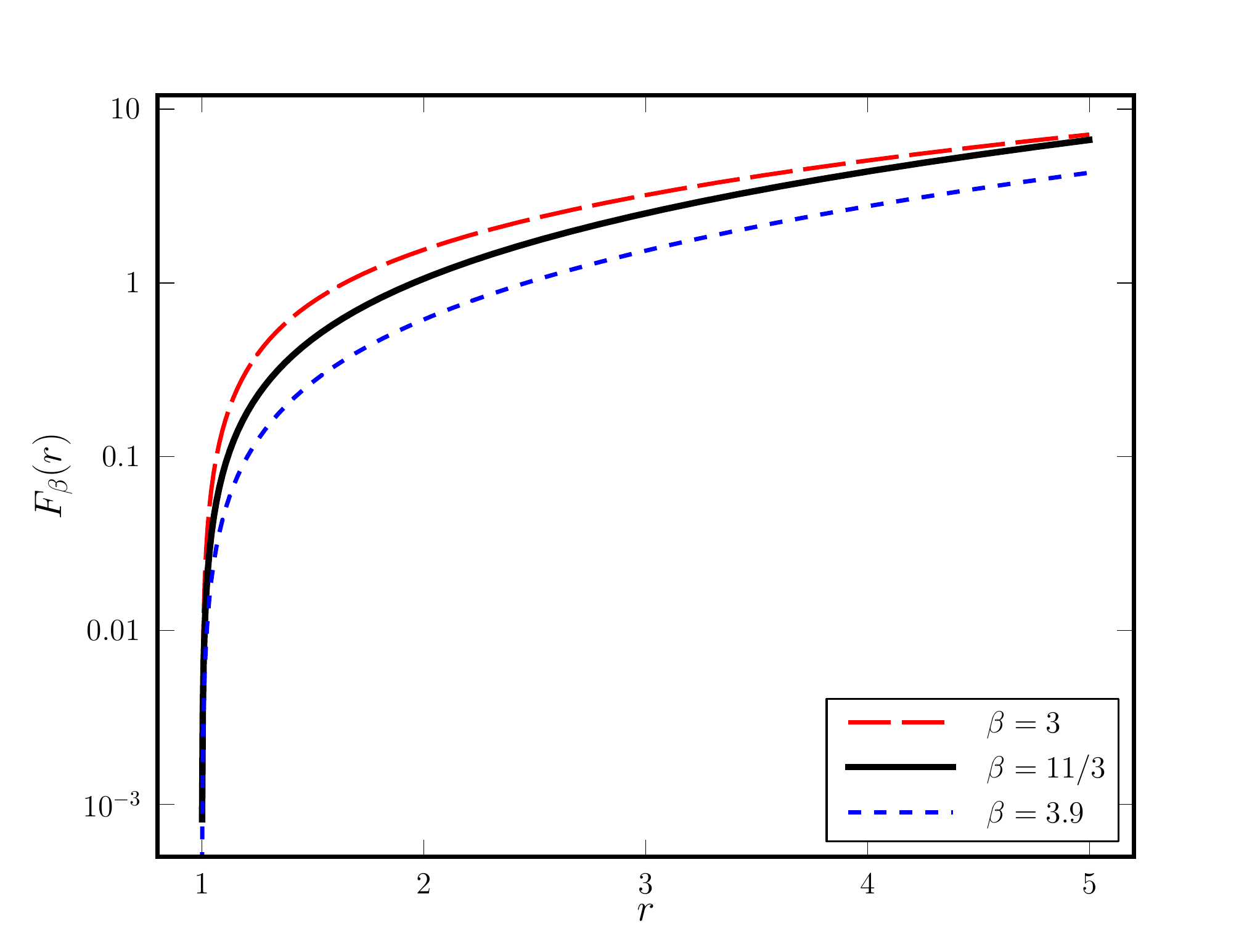}
\hspace {-0.3in}
\includegraphics[scale=0.40]{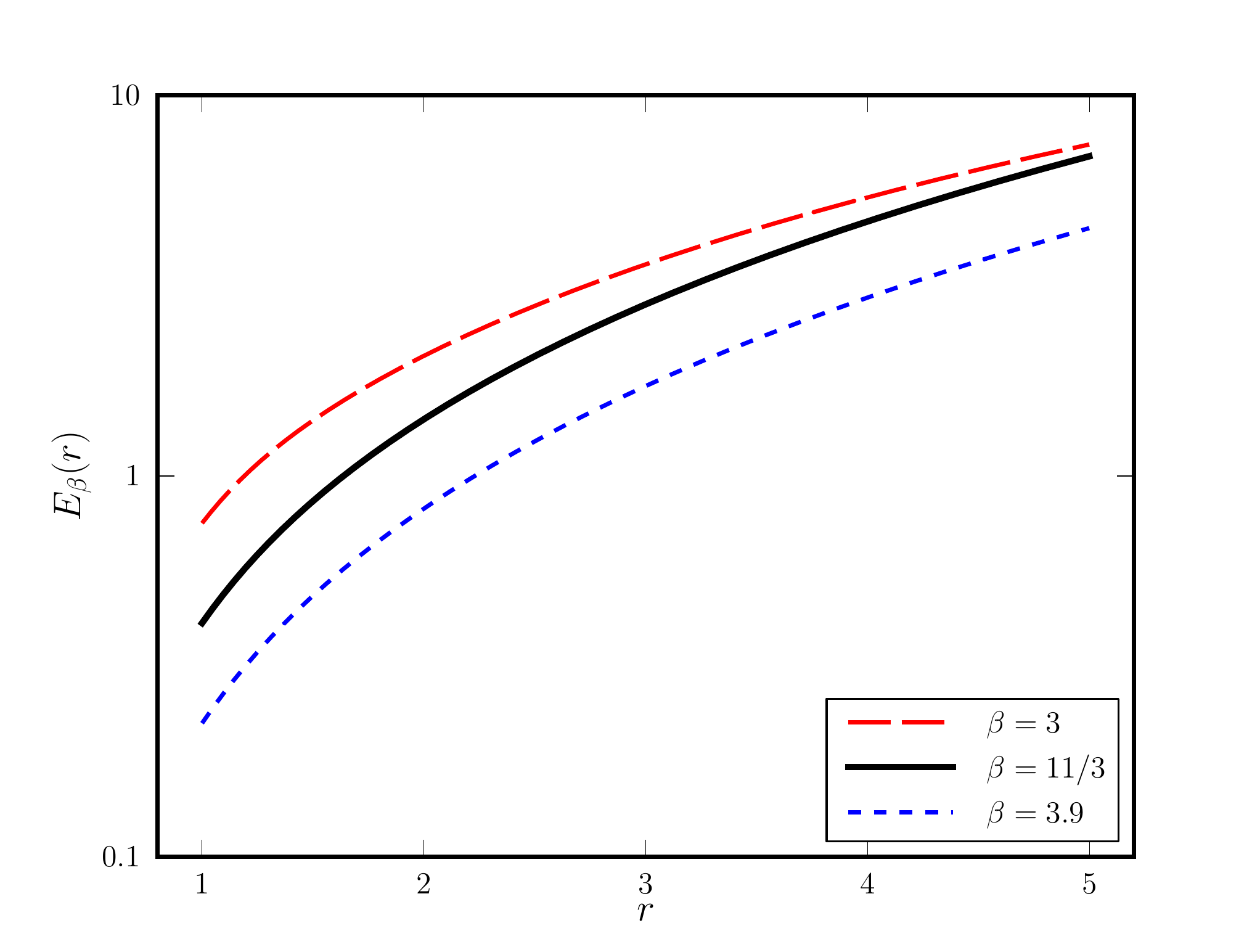}
\caption{\footnotesize 
(Left) The frequency scaling function $F_\beta(r)$ vs.  $r = \nu/\nup$ for three values of $\beta$.
(Right) 
The frequency scaling function $E_\beta(r)$ for the same three values of $\beta$. 
\label{fig:Fbeta}
\label{fig:Ebeta}
}
\end{center}
\end{figure}

We calculate the mean-square of the difference
$\Delta \DMbar(\nuh, \nul, \xvec) = \DMbar(\nul, \xvec) - \DMbar(\nuh, \xvec)$ defined in the main text to
get the two-frequency structure function, 
\be
 \sigmadDMbar^2(\nuh, \nul) &= &
	\left\langle
		\left[
		\Delta \DMbar(\nuh, \nul, \,\xvec)
		\right]^2		  
	\right\rangle
=
 \Cdmbarnn +  \Cdmbarnpnp -2\Cdmbarnnp
\label{eq:app_sf1}
\ee
where the cross correlation of $\delta\DMbar = \DMbar - \left\langle \DM \right\rangle$ between two frequencies is
\be
C_{\delta \DMbar}(\nuh, \nul) = \left\langle \delta\DMbar(\nuh, \xvec) \delta\DMbar(\nul, \xvec) \right\rangle
	=
	 \iint \! d\xvecp d\xvecpp 
	 \iint \! d\zp d\zpp\, 
	\Anuxpz \Anuxppz\langle 
	\delta n_e(\xvecp, \zp) \delta n_e (\xvecpp, \zpp)\rangle.
\ee
The $z$ integrals are from $0$ to $D$ and the $\xvec$ integrals are over an infinite plane.
We define the cross-correlation function of the averaging function, 
\be
C_A(\dxvec, z, \nuh, \nul) =  \int d\xbarvec \, A_{\nul}(\xbar-\dxvec/2, z) A_{\nuh}(\xbar-\dxvec/2, z),
\ee
and assume that it changes slowly in $z$. 
By changing variables from $\xvecp, \xvecpp$ to 
$\xbarvec = (\xvecp + \xvecpp)/2$ and $\dxvec = \xvecp - \xvecpp$
and from $\zp, \zpp$ to $\zbar = (\zp+\zpp)/2$ and $\dz = \zp-\zpp$,
and using the hierarchy of scales assumed in Appendix~\ref{app:scatt},
the integration over $\dz$ gives $\dunit \delta(q_z)$ and we obtain
\be
 \sigmadDMbar^2(\nuh, \nul)
	&=& 
	\frac{1}{2}
	 \int d\zbar \!\! \int \! d\delta\xvec \, \int d\delta z \Dne(\delta\xvec, \delta z;  \zbar) 
	\left[ 
		2 C_A(\delta\xvec, \zbar, \nuh, \nul)
		- C_A(\delta\xvec, \zbar, \nuh, \nuh)
		- C_A(\delta\xvec, \zbar, \nul,\nul)
	\right];
\label{eq:app_sf2}
\ee
$\Dne(\delta\xvec, \delta z;  \zbar) = 
	\left\langle [\delta n_e(\xvec, z)  - \delta n_e(\xvec+\delta\xvec, z+\delta z)   ]^2 \right\rangle$
is the structure function for the electron density. 
For the power-law spectrum of Appendix~\ref{app:scatt}  we have
\be
\int d\delta\zbar \, \Dne(\delta\xvec, \delta z, \zbar) = f_\beta \cnsq(\zbar) \left(\delta x \right)^{\beta-2}.
\ee
We now adopt the Gaussian smoothing function of Eq.~\ref{eq:smooth} and employ the frequency scaling of  
$\sigma_X$, which is the  same as that of the
 RMS scattering angle $\sigtheta \propto \nu^{-x_\theta}$ with $x_{\theta} = \beta/(\beta-2)$.   By referencing frequency-dependent quantities  to the higher frequency $\nuh$ we get 
\be
\sigmadDMbar^2(\nuh, \nul) = 
		\frac{(2\pi)^2 \Gamma(2-\beta/2)} {\beta-2} 
		F^2_\beta(\nuh/\nul) \int d\zbar\, \cnsq(\zbar) \sigma_X^{\beta-2}(\zbar, \nuh)
\label{eq:DDMbar1}
\ee
where the frequency-scaling function is
\be
F_\beta(r) = \left\{ 2^{(4-\beta)/2} \left[ 1 + r^{2\beta/(\beta-2)} \right]^{(\beta-2)/2} - r^{\beta} -1 \right\}^{1/2}.
\label{eq:app_Fbeta}
\ee
Figure~\ref{fig:Fbeta} shows $F_\beta(r)$ along with a related  function 
$E_\beta(r) = r^2 F_\beta(r) / (r^2 - 1)$ that is used in the main text to characterize time-of-arrival errors.

To evaluate the integral in Eq.~\ref{eq:DDMbar1} we need expressions for how the transverse extent of
the ray path bundle varies with $z$.    We define
\be
\sigma_X(\zbar, \nu) = D\sigtheta(\nu) h(z/D)
\label{eq:sigX}
\ee
where $h(x) = 1-x$ for a thin screen, $h(x) = x(1-x)$ for a statistically homogeneous medium,
and $h(x) = 1$ for a plane wave incident on the medium.  We then define
\be
H_\beta = 
	\frac{\int d\zbar \cnsq(\zbar) [h(\zbar/D)]^{\beta-2}}{\int d\zbar \cnsq(\zbar)}
	\equiv \left\langle [h(z/D)]^{\beta-2} \right\rangle_{\cnsq}.
\ee
For a thin screen, $H_\beta = (\Dp/D)^{\beta-2}$, 
for a statistically uniform medium, $H_\beta = \int_0^1 dw\, [w(1-w)]^{\beta-2}$, and for plane-wave
incidence, $H_\beta$ is the limit of the thin-screen case as $\ds \to \infty$. 

Using Eq.~\ref{eq:sigtheta} to evaluate $\sigtheta(\nuh)$, the RMS DM difference becomes
\be
\sigmadDMbar(\nuh, \nul) &=&
	G_\beta
	Q_\beta \, 
	c^{\beta/2} r_e \,
	D^{(\beta-2)/2} \,
	\SM \,	
	\nu^{-\beta/2}F_\beta(\nu/\nup)
\label{eq:DDMbar2}
\ee
where the two dimensionless quantities are
\be
	Q_\beta =  \left[
				\frac{(\sqrt{2}\pi)^{4-\beta} \Gamma(2-\beta/2)   f_\beta   }{(\beta-2) }
			      \right]^{1/2},
 	&&
	\quad\quad\quad	
	G_\beta = \left[ H_\beta \left(\frac{\SMeff}{\SM}\right) \right]^{1/2}.
\label{eq:DDMbar2parameters}
\ee
All of the geometry-dependent factors are consolidated into $G_\beta$.  

We also express $\sigmadDMbar(\nuh, \nul)$ in terms of the Fresnel phase by substituting for $\SM$ from 
Eq.~\ref{eq:SMFres}, 
\be
\sigmadDMbar(\nuh, \nul) &=&
	g_\beta
	q_\beta \, 
	\nu F_\beta(\nu/\nup) \,
	\left( \frac{ \phi_F^2 }{c r_e} \right)
	\\
	\nonumber \\
	q_\beta = 	\left[
				     \frac{2^{\beta/2}\pi^2  \Gamma(2-\beta/2)    }{(\beta-2)  f_\beta}			
			     	\right]^{1/2},	
	&&
	\quad\quad\quad
	g_\beta = \left[
				H_\beta   
				\left(\frac{D}{\Deff}\right)^{\beta-2}
				\left(\frac{\SMeff}{\SM}\right) 
			\right]^{1/2}.
\label{eq:DDMbar3}
\ee
Values of $Q_\beta, G_\beta, q_\beta$ and $g_\beta$ are given in Table~\ref{tab:parameters} for
a Kolmogorov spectrum. 

\begin{deluxetable}{lccc}
\tabletypesize{\Large}
\tablecaption{\label{tab:parameters}Relevant  Factors for a Kolmogorov Medium ($\beta=11/3$)}
\tablecolumns{4}
\startdata
\hline
\\
\multicolumn{2}{l}{\hspace{-0.1in} {\bf Geometry Independent Factors:}}  \\
\\
 Quantity & $\beta = 11/3$ \\
\\
\hline
\\
\quad $f_\beta$ & 				$88.3$	\\			
\\
\quad $Q_\beta$ & 				$22.0$ \\
\\
\quad $q_\beta$ & 				$1.15$\\
\\
\quad $F_\beta(r)$ & 	$F_{11/3}(1) = 0$ & $F_{11/3}(2) = 1.056$ & $F_{11/3}(5) = 6.64 $  \\
\\
\hline
\\
\multicolumn{2}{l}{\hspace{-0.1in} {\bf Geometry Dependent Factors:}}  \\
\\
 Quantity & Thin Screen & Uniform & Plane Wave \\
               & ($x = \ds/D$) \\
 \\
 \hline
\\
\quad $\Deff / D$ &  				$x(1-x)$ 				& 	$1/4$			& 	1	\\
\\
\quad $\SMeff/SM$ & 				$x^{5/3}$				& 	$3/8$			&	1	\\
\\
\quad $h(z)$	  &				$1-z/D$				&	$(z/D)(1-z/D)$		& 	1	\\
\\
\quad $H_\beta$ & 				$(1-x)^{5/3}$			& 	$0.056$			&	$(1-x)^{5/3}$ as $x\to 0$	\\
\\
\quad $G_\beta$ & 				$[x(1-x)]^{5/6}$				&	$0.145$			&	1	\\
\\
\quad $g_\beta$ & 				$1$				&	$0.46$			&  	1	\\
\enddata
\end{deluxetable}

\end{appendix}




\end{document}